\documentclass[twocolumn]{aastex631}

\newcommand{\msun}{$M_{\odot}$}
\newcommand{\sersic}{S\'ersic}
\newcommand{\mkgroup}{$M_{\rm K,\, group}$}
\newcommand{\mstar}{$M_*$}
\newcommand{\satgen}{\texttt{SatGen}}
\usepackage{amsmath} 

\graphicspath{{./}{figures/}}

\begin{document}

\title{ELVES III: Environmental Quenching by Milky Way-Mass Hosts}

\author[0000-0002-5612-3427]{Jenny E. Greene}
\affiliation{Department of Astrophysical Sciences, Princeton University, Princeton, NJ 08544, USA}
\author[0000-0002-1841-2252]{Shany Danieli}
\altaffiliation{NASA Hubble Fellow}
\affiliation{Department of Astrophysical Sciences, Princeton University, Princeton, NJ 08544, USA}
\author{Scott Carlsten}
\affiliation{Department of Astrophysical Sciences, Princeton University, Princeton, NJ 08544, USA}
\author{Rachael Beaton}
\affiliation{Space Telescope Science Institute, Baltimore, MD, 21218, USA}
\affiliation{Department of Astrophysical Sciences, Princeton University, Princeton, NJ 08544, USA}
\author{Fangzhou Jiang}
\affiliation{Carnegie Observatories, 813 Santa Barbara Street, Pasadena, CA 91101, USA}
\affiliation{TAPIR, California Institute of Technology, Pasadena, CA 91125, USA}
\author[0000-0001-9592-4190]{Jiaxuan Li}
\affiliation{Department of Astrophysical Sciences, Princeton University, Princeton, NJ 08544, USA}



\begin{abstract}

Isolated dwarf galaxies usually exhibit robust star formation but satellite dwarf galaxies are often devoid of young stars, even in Milky Way-mass groups. Dwarf galaxies thus offer an important laboratory of the environmental processes that cease star formation. We explore the balance of quiescent and star-forming galaxies (quenched fractions) for a sample of $\sim 400$ satellite galaxies around 30 Local Volume hosts from the Exploration of Local VolumE Satellites (ELVES) Survey. We present quenched fractions as a function of satellite stellar mass, projected radius, and host halo mass, to conclude that overall, the quenched fractions are similar to the Milky Way, dropping below 50\% at satellite \mstar$\approx 10^8$~\msun. We may see hints that quenching is less efficient at larger radius. Through comparison with the semi-analytic modeling code \satgen, we are also able to infer average quenching times as a function of satellite mass in host halo-mass bins. There is a gradual increase in quenching time with satellite stellar mass rather than the abrupt change from rapid to slow quenching that has been inferred for the Milky Way. We also generally infer longer average quenching times than recent hydrodynamical simulations. Our results are consistent with models that suggest a wide range of quenching times are possible via ram-pressure stripping, depending on the clumpiness of the circumgalactic medium, the orbits of the satellites, and the degree of earlier preprocessing.

\end{abstract}

\section{Introduction}

Low-mass (dwarf) galaxies have become a critical testing ground for cold dark matter on small scales \citep[e.g.,][]{Bullock:2017,Sales:2022}. Interconnected with the desire to use them as cosmological probes is the need to understand the baryonic processes that shape dwarfs. In particular, dwarfs are an especially sensitive probe of the role of environment in shaping their star-formation history. When left to their own devices, dwarf galaxies (particularly with $M_* \approx 10^8-10^{10}$~\msun) are star-forming $\sim 95\%$ of the time \citep{Geha_2012}, although exceptions are known \citep[e.g.,][]{Polzin:2021aa}. The fact that most satellites of the Milky Way (MW) are quiescent \citep[or ``quenched'', e.g.,][]{Hodge:1971} suggests that becoming a satellite drastically changes the gas content and star-forming capacity of dwarfs. Indeed, this gas deficiency has long been directly observed as a function of distance from the MW and M31 \citep{Einasto:1974,Grcevich_2009,Spekkens_2014,Putman_2021,Karunakaran:2022gas}. 

Theoretically, there are a number of processes thought to drive environmental changes in dwarf galaxy structure and star formation history. Tides from the host halo can lead to stripping (removal of dark matter and stars as the satellite passes through the host potential) and/or stirring \citep[in which tides impart energy to the satellite and change the dynamics of stars; e.g.,][]{Mayer:2001,Kravtsov:2004,Mayer:2006,Diemand:2007,Kazantzidis:2017}. These tidal interactions may be particularly dramatic between satellites and host galaxy disks \citep[e.g.,][]{Donghia:2010} or for satellites on very radial orbits \citep[e.g.,][]{Mateo:2008}. Galaxies may slowly quench by running out of fresh gas \citep{Larson:1980}. Stellar feedback can also contribute to gas loss \citep[e.g.,][]{Agertz:2013}. However, specifically in low-mass galaxies, ram-pressure stripping is likely the most important factor in removing gas from the galaxy \citep[e.g.,][]{Grebel:2003,vanGorkum:2004,Tonnesen:2009}. Ram-pressure from the gas in the parent halo can strip the gas from an incoming satellite \citep{GunnGott:1972}, and \citet{LinFaber:1983} already suggested that a hot gaseous halo might be responsible for the smooth and red morphology of dwarf galaxies orbiting the MW \citep[][]{Mayer:2006,Grebel:2003}. The inferred circumgalactic medium densities required to reproduce the quenched fractions in the MW appear reasonable \citep{Grcevich_2009} and hydrodynamical simulations succeed at reproducing the quenched fraction of satellites as a function of their mass as observed in the MW and M31 \citep[e.g.,][]{Simpson_2018,Akins_2021,Samuel_2022,Font_2022}. 

On the other hand, quenching is unlikely to proceed uniformly given that the range of densities and clumpiness in the circumgalactic media \citep{Simons:2020} and the detailed orbits of individual satellites are likely to matter \citep{Emerick_2016,Fillingham:2019}. At the same time, there is evidence that a circumgroup medium of the MW combined with M31 may be more effective at quenching the MW satellites \citep{McConnache:2007,Garrison-Kimmel:2019,Putman_2021}. It would be valuable to have a larger sample of satellites to empirically test quenching timescales as a function of satellite and host properties.

Luckily, such group catalogs of MW-mass halos with completeness to classical satellite masses (e.g., $M_* \sim 10^5-10^6$~\msun) have recently become available \citep{Merritt:2014,Karachentsev:2015,Geha_2017,Danieli:2017,Bennet:2017,Smercina_2018,Bennet:2019,Crnojevic:2019,Muller:2019,Danieli:2020,Garling:2021,Mao_2021,Mutlu-Pakdil:2022}. In this paper, we focus on the 30 groups within the Local Volume ($D<12$~Mpc) compiled by the Exploration of Local VolumE Satellites Survey (ELVES hereafter). These groups are effectively a volume-limited sample, with hosts comparable to or more massive than the MW ($-22.3 > M_K > -24.9$ mag), and complete in satellite mass to $M_V \approx -9$ \citep[][C22 hereafter]{Carlsten:2022survey}. Surface brightness fluctuation distances \citep{Tully:1987aa}, as calibrated for dwarfs \citep{Carlsten:2019calib}, allow for rapid progress in distance determination for a large number of satellites. 

As shown by C22, at the most basic level, the fraction of galaxies that are quenched as a function of satellite mass matches the measurements for the MW and M31. However, there is more information in the ELVES sample to investigate, such as how the quenched fraction changes with radial distance from the host, with the halo mass of the host, and even potentially with the morphology of the host. We investigate these detailed questions here. We also leverage a custom stellar-to-halo mass relation measured for the ELVES sample by \citet{Danieli:2022}, and a set of custom simulations run with the semi-analytic model \satgen\ \citep{Jiang_2021_satgen} to derive characteristic quenching times based on theoretical infall-time distributions.

The paper begins with a short review of the C22 sample in \S \ref{sec:elves}, followed by an empirical look at how quenched fractions vary with host and satellite properties (\S \ref{sec:fractions}). We then use \satgen\ models to translate quenched fractions into average quenching times (\S \ref{sec:satgen}) and finally discuss the implications for quenching models (\S \ref{sec:discussion}). Throughout we assume a concordance $\Lambda$CDM cosmology here, with $H_0=70$~km s$^{-1}$ Mpc$^{-1}$, $\Omega_{\rm M} = 0.3$, $\Omega_{\Lambda} = 0.7$.

\section{ELVES Survey}
\label{sec:elves}

We briefly summarize the ELVES survey, but refer to C22 for a full description of the survey, to \citet{Carlsten:2021LF} for the satellite luminosity functions, to \citet{Carlsten:2022gc} and \citet{Carlsten:2021structure} for a look at the star cluster populations and satellite galaxy structures, respectively, and to \citet{Carlsten:2020survey} for a more in-depth discussion of the satellite search algorithm.

\subsection{Host Galaxy Selection}

ELVES is a volume-limited survey out to $D < 12$~Mpc, the limit of our ability to measure reliable surface brightness fluctuations from the ground. Hosts are selected primarily to have stellar masses similar to the Milky Way. In practice, we employ a $K-$band magnitude lower limit of $M_{\rm K,s} < -22.1$~mag, which roughly translates to a stellar mass limit of \mstar$\approx 10^{10}$~\msun\ \citep{Mcgaugh:2014}. Along with the magnitude and distance cut, we excise regions of the sky dominated by our Galaxy with a Galactic latitude cut: $\lvert b \rvert < 17\fdg4$. This particular cut was tuned to include NGC~891 and Centaurus A, in order to exploit extensive prior work cataloging the groups in each case \citep[e.g.,][]{Crnojevic:2019,Muller:2019,Trentham:2001aa,Muller:2021}. There are a few cases of galaxies that obey these selection criteria, but are projected to fall within the halo of a more massive galaxy (e.g., M82 is in the M81 group). We have used the catalog from \citet{Kourkchi:2017} to make group determinations in individual cases based on the projected distance (within the second turn-around radius or $\sim$ virial radius $R_{\rm vir}$) and radial velocity (less than twice the velocity dispersion of the group). The full list of massive satellites that otherwise obey the ELVES cuts is found in Appendix A of C22. This leaves a sample of 31 potential hosts, of which only NGC~3621 has not been surveyed, for a total of 30 ELVES hosts.

For context, the Satellites Around Galactic Analogs \citep[SAGA;][]{Geha_2017,Mao_2021} Survey uses an $M_{K}-$based selection to identify Milky Way analogs, including a high- and low-luminosity cut. In contrast, ELVES does not have a bright limit for hosts, but rather has a volume limit that includes all hosts more luminous than our limit. In practice, this means that ELVES includes some groups that are at least an order of magnitude more massive than the Milky Way (e.g., the M81 group). 

\subsection{Satellite Selection and Completeness}

Of the 30 ELVES hosts, five (MW, M31, Cen A, NGC 5236, and M81) had cataloged satellites to a comparable or deeper satellite mass limit before ELVES. In the 25 remaining hosts, the candidate satellites are identified by C22, and then distances are determined either from literature measurements (either redshift or tip of the red giant branch) or from surface brightness fluctuations. We describe satellite candidate selection followed by distance determination in the following subsection.

We first must identify candidate satellites in an imaging field. C22 uses two primary sources of imaging. Six hosts (NGC~1023, NGC~4258, NGC~4565, NGC~4631, NGC~5194, and M104) were surveyed in \citet{Carlsten:2020survey} using Canada-France-Hawai'i Telescope/MegaCam imaging. The remaining seventeen hosts (as well as the outer parts of NGC~4631 and NGC~4258) use imaging from the Dark Energy Camera Legacy Survey \citep[DECaLs;][]{Zou:2017,Zou:2018,Dey:2019aa}. Our goal was to achieve coverage to $R=300$~kpc (roughly the virial radius of the MW) but that was not achieved for all hosts.

For this purpose, C22 utilizes custom software that identifies and masks stars \citep{Gaia:2018} and bright galaxies, and then searches for statistically significant groups of pixels above the background \citep[for more discussion of the methodology, see also][]{Greco:2018aa}. {\tt SExtractor} \citep{Bertin:1996aa} is run to detect sources with both a high ($15\sigma$) and low ($1\sigma$) threshold, in order to mask not only the bright cores of massive, background galaxies but their low surface brightness outskirts, which can be a major contaminant in searches for faint and fuzzy galaxies. Then the masked image is filtered with a Gaussian kernel $\sim 2 \times$ the point-spread function to increase the contrast for faint groups of pixels. A second detection is run with a significance threshold of $2 \sigma$ and size of at least 4\arcsec\ ( or 800 pixels, as implemented by the minimum area parameter in SExtractor). The last step is a visual inspection that removes artifacts such as saturation spikes, galaxy outskirts, and clear background galaxies \citep[e.g., small sources with clear spiral structure; see ][]{Carlsten:2020survey}. Detection is run in $g$ and $r$ or $i$ independently, and then the merged catalog only includes objects detected in both bands. 

Survey completeness is calculated with artificial galaxy injection tests. Galaxies are modeled as single \sersic\ \citep{Sersic:1968aa} profiles with $n=1$ (exponential profiles), which \citet{Carlsten:2021structure} show provide a good description of the light profile (albeit real galaxies prefer $n \approx 0.7$). The galaxy magnitudes cover the full range on our sample and colors are likewise drawn from a range $0.3 \leqslant g-r \leqslant 0.6$, motivated by the real data. Then the automated parts of the detection algorithm are run, and completeness is computed as a function of magnitude and size. We are $\sim 50\%$ complete at $M_V < -9$~mag and $\mu_{0, V} < 26.5$~mag~arcsec$^{-2}$. This magnitude corresponds roughly to $M_* \approx 10^{5.5}$~\msun\ for an average color of $\langle g-i \rangle = 0.7$ mag. 

\subsection{Distances and Final Catalog}

With candidate satellites in hand, we must then determine distances to all satellite candidates. The DECaLS data have neither the depth nor the spatial resolution to enable robust surface brightness fluctuation measurements, and so distances in general are determined with follow-up imaging from Gemini, Magellan, or archival Hyper Suprime-Cam images. ELVES relies heavily on surface brightness fluctuation distances \citep{Tonry:1988} to build up such a large sample of dwarf satellites without giant investments with the \emph{Hubble Space Telescope} \citep[e.g.,][]{Danieli:2017} or spectroscopic follow-up as in SAGA \citep{Geha_2017,Mao_2021}. 

Surface brightness fluctations (SBF) are PSF-scale fluctuations in star counts, and these fluctuations decrease with increasing distance as more and more stars fall within a single PSF. SBF works because most of the light comes from a small number of the most luminous stars, allowing us to probe their Poisson fluctuations \citep[e.g.,][]{Jensen:1998,Blakeslee:2009}. However, we therefore need to understand the absolute number of massive stars in a population to utilize SBF; typically the SBF magnitude is empirically calibrated to galaxy color, which is a proxy for the stellar populations of a galaxy \citep[e.g.,][]{Cantiello:2018}. \citet{Carlsten:2019calib} present a direct calibration of SBF magnitudes using distances to dwarf galaxies derived from tip of the red giant branch measurements and show that even in dwarf galaxies, where there is ongoing star formation and a wide range of recent star formation histories, we achieve 15\% distance accuracy with SBF \citep[see also][]{Greco:2021}. At 7~Mpc, satellites within 1~Mpc of the host will be scattered into our survey given the distance errors, which \citet[][]{Carlsten:2020radial} estimate provides $\lesssim 10\%$ contamination. The distance limit of ELVES represents the approximate limit of ground-based surface-brightness fluctuation methods to determine distances. 

Satellite candidates will ultimately fall into one of the following groups. Bright satellites have an existing literature distance from tip of the red giant branch measurements or redshift. All other satellites are either confirmed to be at the distance of the host, confirmed to lie in the background, or too faint or low surface brightness to be confirmed. In the end, over all 30 hosts, there are 553 new candidates discovered by ELVES, and an additional 87 satellites from the literature. Of these, 136 are SBF-confirmed, 202 are confirmed in other ways, 172 are shown to sit in the background, 24 are unusable (e.g., due to a nearby bright star), and 106 are possible satellites whose distances are not confirmed. 

Of the 106 ``possible'' satellites, each is assigned a weight, which is the probability that it is a real satellite. Each confirmed or rejected satellite is placed in the $M_V-\mu_{0, V}$ plane. Confirmed satellites fall in a well-defined sequence in this plane due to the mass-size relation, while contaminants, being nearly exclusively in the background, are far more likely to be more compact at a given $M_V$. Thus, we can empirically determine the probability that an unconfirmed satellite is at the host distance based on the location in this plane. In practice, the weight is calculated from the 20 confirmed and background galaxies that are nearest in this $M_V-\mu_{0, V}$ plane. The fraction of confirmed galaxies among those 20 is the weight. We should note that very compact galaxies could be erroneously assigned to the background. On the other hand, as shown by \citet{Carlsten:2020survey} we are not complete for such galaxies. All distance-confirmed satellites have $P_{\rm sat} = 1$. As our default sample, we will consider all potential satellites weighted by $P_{\rm sat}$. The typical satellites fall between our mass limit and $\sim 10^8$~\msun\ in stellar mass, but the richest hosts have satellites extending to masses greater than $10^{10}$~\msun\ (Fig. \ref{fig:sattwod}).

As mentioned above, ELVES does not have uniform radial coverage for all hosts. Of the 30 hosts, there are 21 hosts with projected coverage to $R_p=300$~kpc, and five more with coverage to $R_p=200$ kpc. In this paper, we focus on these 26 hosts with maximum coverage $R_p\geq 200$ kpc, eliminating the four ELVES hosts where we have coverage only to $R_p = 150$~kpc. The other demographic measurement from C22 that we will utilize extensively here is \mkgroup. This is the total $K-$band magnitude of the group, including the host and companions. For a large fraction of the sample, the host completely dominates this value, but for the most massive groups, \mkgroup\ is significantly more luminous due to massive satellites. Thus, \mkgroup\ is our most robust proxy for the dark matter halo mass. In general, to avoid propagating forward uncertainty in $M_h$ and $R_{\rm vir}$, we will work in kpc space, although we check that our results would hold in $R_{\rm vir}$ coordinates. The one exception is when we look for radial trends in quenching across the full sample in \S \ref{sec:fractions_radial}. We note that when we limit attention to the ELVES sample with coverage out to $200$~kpc, we cover $0.8-1 R_{\rm vir}$ for all hosts (and well beyond $R_{\rm vir}$ for most). However, we systematically cover less than $R_{\rm vir}$ for the higher \mkgroup\ sample. 

In Table \ref{tab:samples}, we summarize the number of hosts and satellites associated with each \mkgroup\ bin (see also Figure \ref{fig:fqmk}), where $N_{\rm secure}$ are the number of confirmed ($P_{\rm sat} = 1$) satellites and the rest are possible satellites. Throughout the paper, in measuring quenched fractions, we only consider galaxies falling within $R_p<250$~kpc from the host, but note that our results would not change for $R=200-300$~kpc.

\subsection{Defining the quenched sample}

In the Milky Way, quenched galaxies may be identified in a variety of ways. One might sensibly define quenched galaxies as those with no \ion{H}{1} stores available for star formation, and there are deep limits on \ion{H}{1} masses for Milky Way satellites \citep[e.g.,][]{Grcevich_2009,Spekkens_2014,Putman_2021}. Alternatively, one might want to define a quenched galaxy as having stellar populations older than some age or having finished star formation more than some time ago. For Milky Way galaxies, we have deep color-magnitude diagrams that allow for very tight constraints on star formation histories \citep[e.g.,][]{Mateo:1998aa,Weisz_2011,McConnachie:2012aa,Weisz:2014,Brown:2014,Skillman_2017}. Finally, one might instead look for more direct signs of ongoing star formation, say from UV or H$\alpha$ emission. SAGA, for instance, defines a star-forming galaxy as one for which they detect H$\alpha$ \citep{Geha_2017}, while \citet{Karunakaran_2021} validate these values in SAGA with \emph{GALEX} data. 

In ELVES, none of these ideal tracers is available for the full sample. \citet{Karunakaran:2022gas} \citep[see also][]{Karunakaran:2020aa} presents some archival and some new \ion{H}{1} constraints to ELVES, but they are not comprehensive. Many galaxies in ELVES have archival \emph{GALEX} and/or H$\alpha$ measurements, but by no means all. Finally, amassing the deep color-magnitude information required to determine star formation histories would be prohibitive with \emph{HST} beyond the 3-4 Mpc that has already been surveyed \citep[e.g.,][]{Dalcanton_2009,Weisz_2011}. 

Thus, we rely instead on galaxy color and morphology. In C22 we present a morphology-based classification for each galaxy, determined by visual inspection. Red and smooth galaxies are classified as early-type, while blue and asymmetrical/lumpy galaxies showing visible signs of star formation are classified as late-type. C22 shows that if we adopt a luminosity-dependent color cut to divide galaxies into early and late-type, we recover very similar quenched fractions. More specifically, \citet{Carlsten:2021structure} used a dividing line of $g-i = -0.067 \times M_V-0.23$, which also seems to work well for the simulations of \citet{Font_2022} and \citet{Pan:2022}. Likewise adopting a mass-dependent color cut in NUV$-g$ yields a virtually identical early/late-type demarcation. It will be interesting to explore in future work how the color-magnitude relation relates to the known mass-metallicity relation for dwarfs \citep{Kirby:2013}. We will revisit this tilt in the color-mass relation in \S \ref{sec:sagacomp}. 


We can go beyond C22 and use both archival H$\alpha$ and published \ion{H}{1} measurements to further support the fidelity of the early/late-type demarcation. \citet{Karunakaran:2022gas} find that within this archival sample, all the \ion{H}{1} detected galaxies are classified as late-type by C22. 
Turning to H$\alpha$, we use the catalog of \citet{Karachentsev:2021}. Most of these measurements come from narrow-band imaging \citep{Kennicutt:2008,Kaisin:2019}. Focusing on ELVES satellites within a range of \mstar$=10^7-10^8$~\msun, there are 39 matches, of which 20 are late-type and all are detected. Of the remaining 19 early-type satellites, 7 have only upper limits in the catalog, reaching star formation rates as low as $<10^{-6}$~\msun~yr$^{-1}$. In the Appendix, we show that the distribution of star-formation rates, as inferred from H$\alpha$, is disjoint between the early-type and late-type galaxy samples. Since we have controlled for stellar mass, this tells us that either we are seeing only a tiny vestige of ongoing star formation in the early-type galaxies, or that the H$\alpha$ arises from some other physical process. Therefore, to be maximally inclusive of all ELVES satellites with and without H$\alpha$, we classify quenched dwarfs as those with early-type designation in C22.

\section{Quenched fractions}
\label{sec:fractions}

\begin{deluxetable}{ccccc}
\tablenum{1}
\tablecolumns{5}
\tablecaption{Satellite Samples\label{tab:samples}}
\tablewidth{0pt}
\tablehead{
 \colhead{Host Sample} &\colhead{Selection} & \colhead{$N_{\rm host}$} & \colhead{$N_{\rm sat}$} & 
\colhead{$N_{\rm secure}$}} 
\decimalcolnumbers
\startdata
High & $-23.5 < M_K < -24.5$ & 4 & 142 & 108 \\
Medium & $-22.5 < M_K < -23.5$ & 13 & 167 & 138 \\
Low & $-21.5 < M_K < -22.5$ & 9 & 99 & 67 \\
All & \nodata & 26 & 408 & 313 \\
\enddata
\tablecomments{Host and satellite sub-samples used in this work, always within 250 projected kpc from the host. (1) Host bins by \mkgroup, our proxy for halo mass. (2) \mkgroup\ range per bin. (3) Number of hosts per bin. The most massive bin contains small groups like M81, that may have had very different evolutionary histories and have satellites of comparable mass to some primaries in the Low bin. (4) Total number of satellites, including those with $P_{\rm sat} < 1$. (5) Confirmed satellites with robust distances ($P_{\rm sat} = 1$).}
\end{deluxetable}

In this section, we empirically quantify how the quenched fraction varies with (a) the host halo mass (proxied by \mkgroup), (b) morphology of the host, (c) the satellite stellar mass, and (d) radial position within the host (in projection, $R_p$). We define the total number of satellites as the sum over all satellites probabilities $P_{\rm sat}$ down to our mass limit. The quenched fraction is then the ratio of the sum over $P_{\rm sat}$ of those satellites that are visually classified as early-type with the sum over $P_{\rm sat}$ for all satellites, within a specified mass and $R_p$ range. However, if we only include galaxies with high $P_{\rm sat} > 0.8$, the results are very consistent with what we present here (as also argued by C22). 

\begin{figure}
\hskip 0mm
\includegraphics[width=0.45\textwidth]{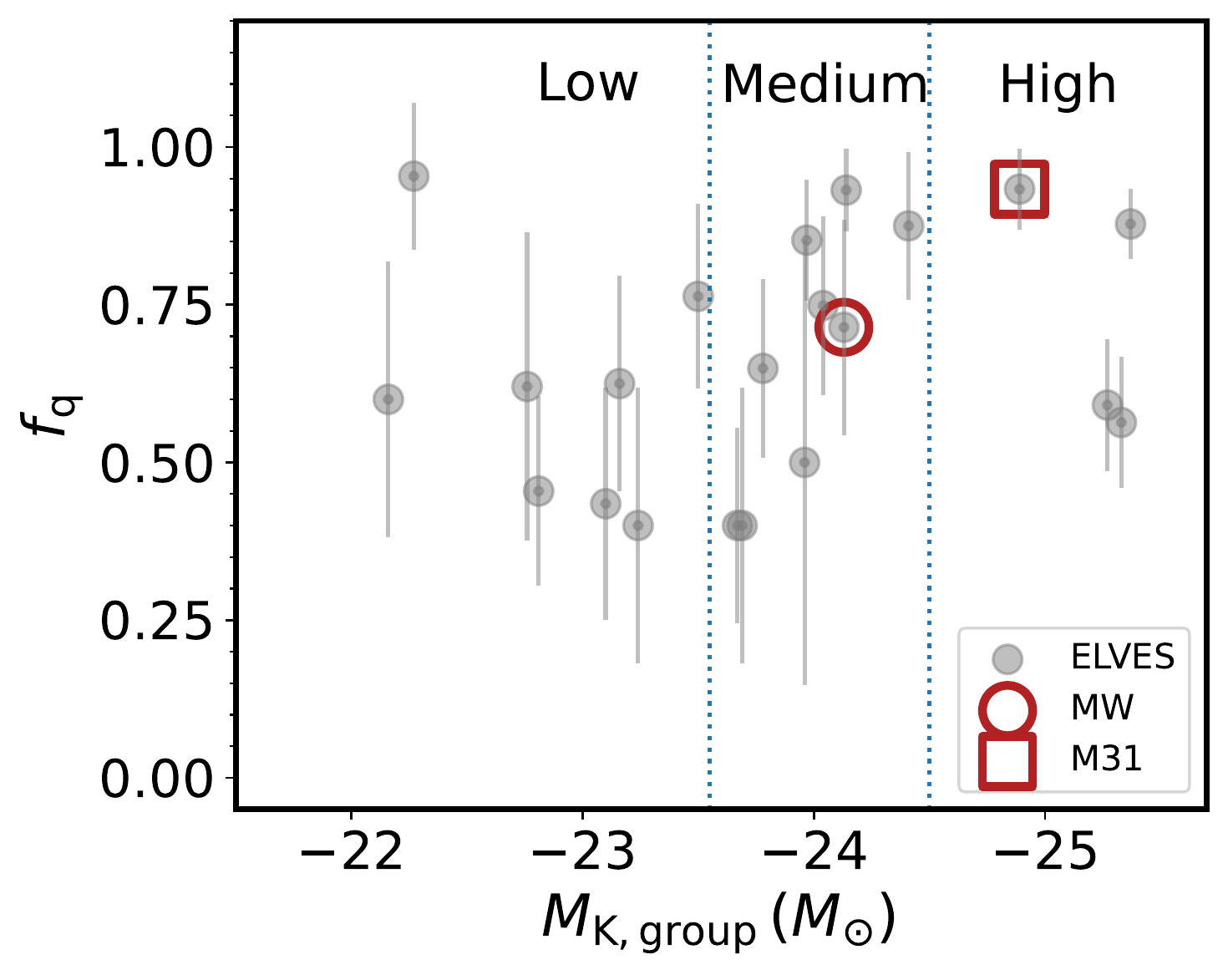}
\caption{The quenched fraction per galaxy for distance to host $R_p<250$~kpc, including all satellites using the satellite probability. Error bars shown here are purely based in the binomial theorem, but in nearly all cases encompass the total spread in $f_q$ based on including or excluding satellites with $P_{\rm sat} < 1$. The correlation between $f_q$ and \mkgroup\ is not significant, given the large scatter in values per galaxy. However, for the purpose of analysis further on in the paper, we do divide the host sample into three bins of \mkgroup\ (high, medium, low) as indicated here. 
\label{fig:fqmk}}
\end{figure}

\begin{figure*}
\includegraphics[width=0.33\textwidth]{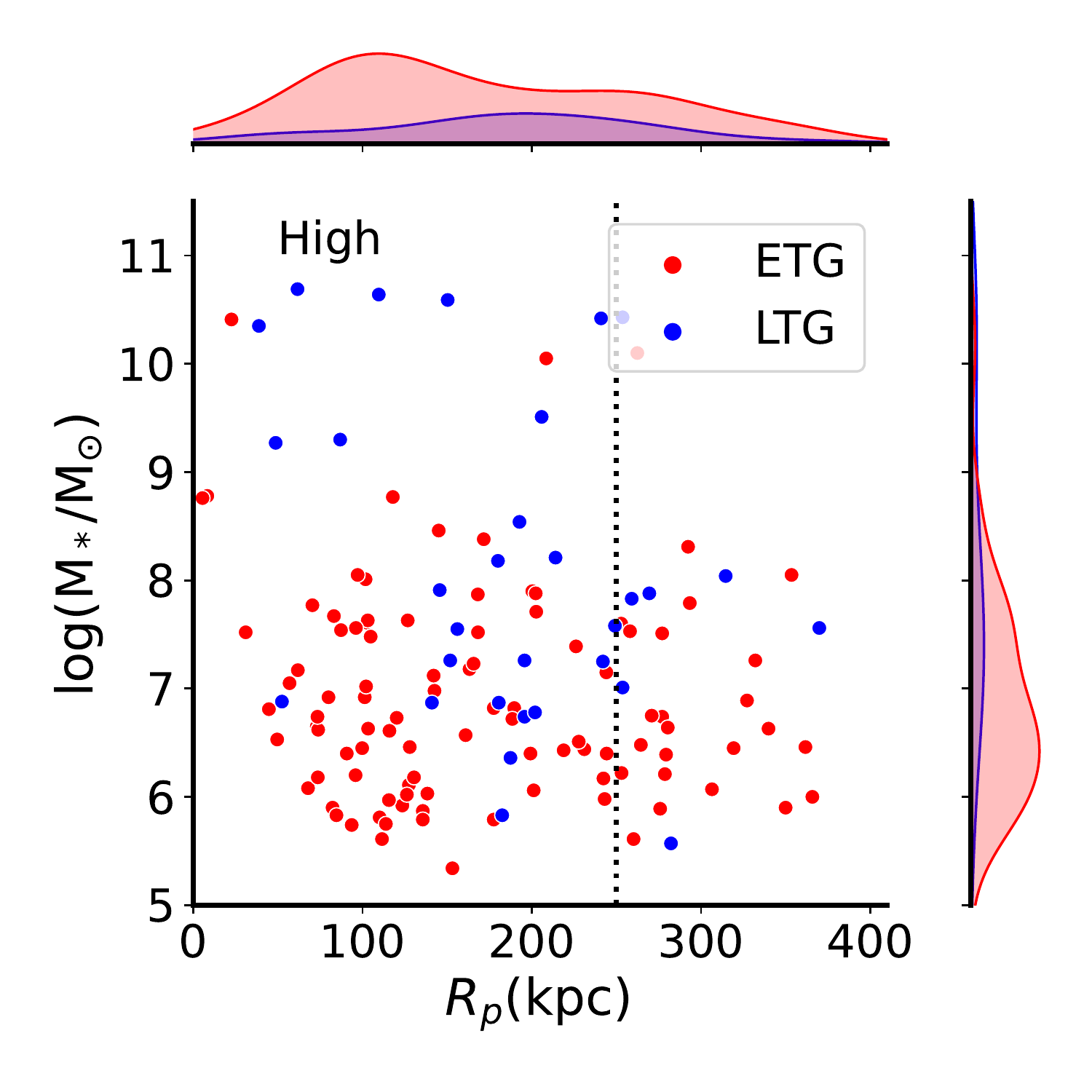}
\includegraphics[width=0.33\textwidth]{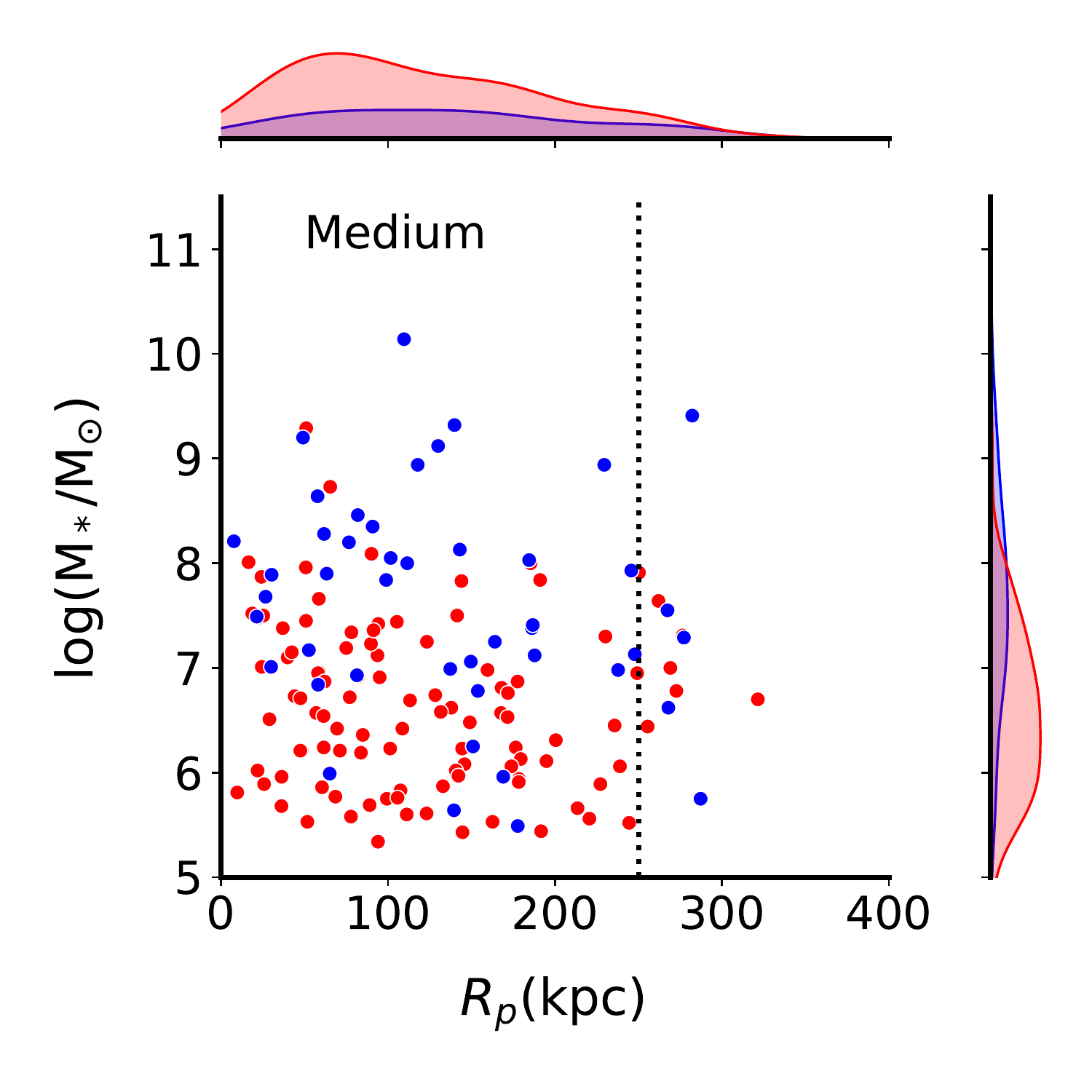}
\includegraphics[width=0.33\textwidth]{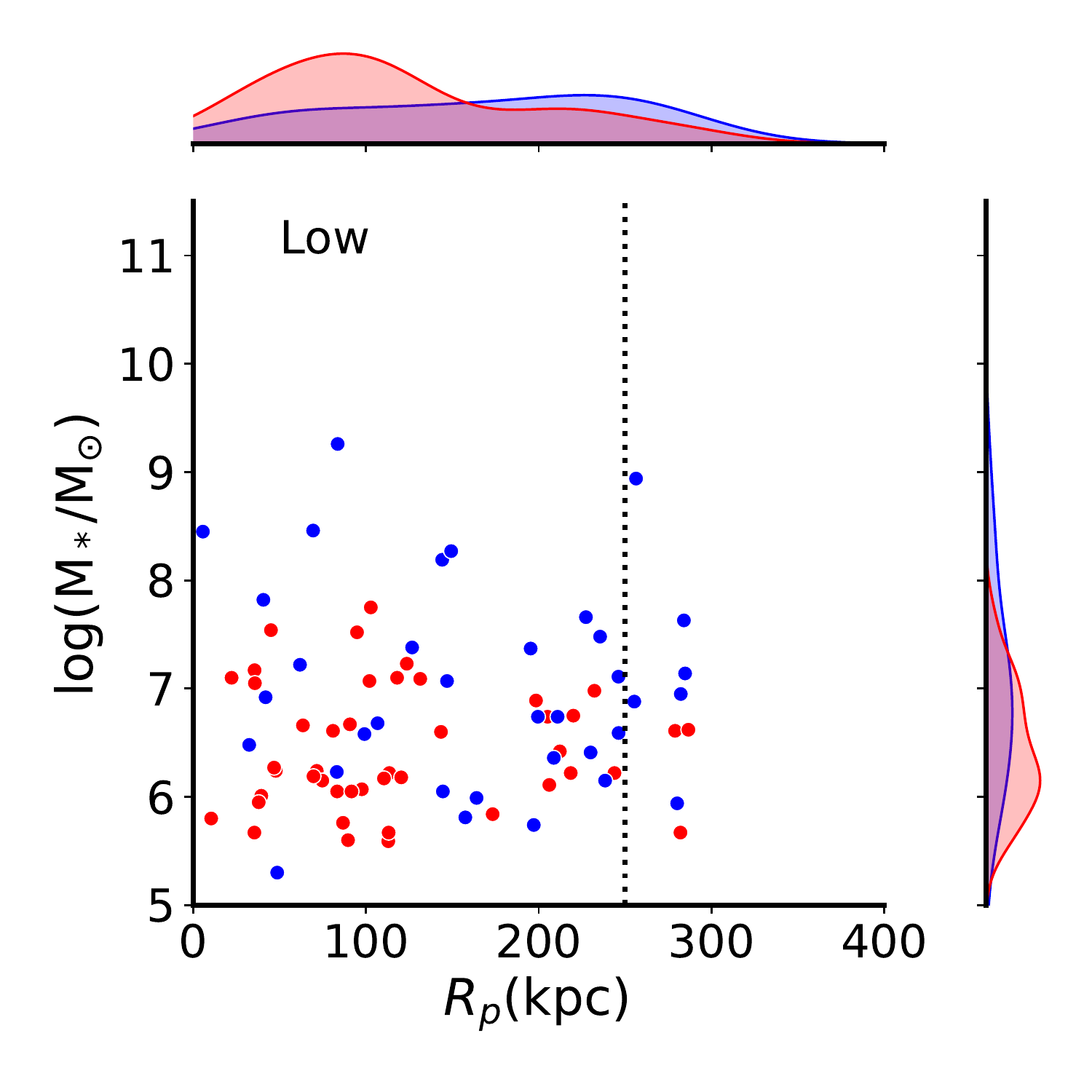}
\caption{ELVES satellites with a high probability to be a satellite ($P_{\rm sat} > 0.5$) as a function of distance from their host, divided into three \mkgroup\ bins, as defined in Figure \ref{fig:fqmk}. Dotted lines at $R_p = 250$ kpc projected from the host indicate the samples used in this work to compute quenched fractions ($f_q$). We distinguish between ETG (red) and LTG (blue), and show the marginal distributions in mass and radius. We clearly see that more massive hosts have more massive satellites as well as trends in quenching as a function of mass and radius. 
\label{fig:sattwod}}
\end{figure*}
\vskip 10mm

\subsection{Individual Quenched Fractions}

We start by asking whether the quenched fractions per galaxy may be correlated with any other properties of the system. In group- and cluster-mass halos, a clear morphology-density relation is observed \citep[e.g.,][]{Dressler:1980,Postman:1984,Zabludoff:1998,Sales:2013}. Our hosts are in much smaller halos, but we do span 
at least an order of magnitude in halo mass (C22). Thus, we can explore whether quenching efficiency varies with host stellar mass or halo mass for the ELVES sample, at least down to a satellite stellar mass of $\sim 5 \times 10^5$~\msun. 

The quenched fraction per host, as a function of \mkgroup, is shown in Figure \ref{fig:fqmk}, measured within $R_P < 250$ kpc for spatial uniformity across the sample. Since many hosts (particularly at the low \mkgroup\ end) have a small number of satellites, the errors on individual quenched fractions can be quite large. Our errors are based purely on the binomial theorem, but we note that in nearly all cases, these errors are larger than the systematic error calculated by including or excluding satellites with $P_{\rm sat} < 1$. Even given these uncertainties, we observe a large spread in the quenched fractions, particularly in the MW-mass range where our host statistics are best. This spread cannot be explained by the uncertainties, and so we investigate whether \mkgroup, host stellar mass, or satellite richness correlates with $f_q$ per host.

We do not find compelling correlations between host properties and $f_q$. Quantitatively, using a Pearson correlation test, the correlation between $f_q$ and \mkgroup\ is not significant (with a 26\% probability of the null correlation).  We also investigate the quenched fraction as a function of host stellar mass and satellite richness. We see no significant correlation with $M_*$ ($P_{\rm null} = 0.1$), nor number of satellites ($P_{\rm null} = 0.2$). It is possible that other properties, like the mass of the stellar halo or other proxies for the accretion histories, may correlate with the quenched fraction, but these are beyond our ability to test at this time.

Interestingly, C22 do report a correlation between \mkgroup\ and $f_q$. However, their quenched fraction is measured only for satellites with \mstar$< 3 \times 10^9$~\msun. In \S \ref{sec:massdist} we will investigate whether the quenched fraction has a dependence on halo mass at fixed satellite mass. Even if present, this dependence will be washed out by the competing change in satellite mass function, whereby more massive halos have more massive satellites, that are preferentially unquenched. One caveat we should mention is that we probe to different fractions of $R_{\rm vir}$ across the sample, which also may muddy the comparison. We directly investigate the relationship between quenched fraction and satellite mass in \S \ref{sec:massdist}. 

Finally, as pointed out by C22, the quenched fraction does not fall to zero in our lowest-mass halos, but seems to flatten at $f_q \approx 50$\%. We caution that at the very lowest \mkgroup, NGC~3344 and NGC~4517 have smaller numbers of satellites ($\Sigma P_{\rm sat} \lesssim 5$ for $r<250$~kpc) and thus the quenched fractions are also uncertain. It will be useful to build larger samples of groups at lower mass, but because the number of satellites per host drops, large samples of hosts will likely be needed. 

\subsection{Quenching and Satellite Mass}
\label{sec:massdist}

\begin{figure*}
\includegraphics[width=0.48\textwidth]{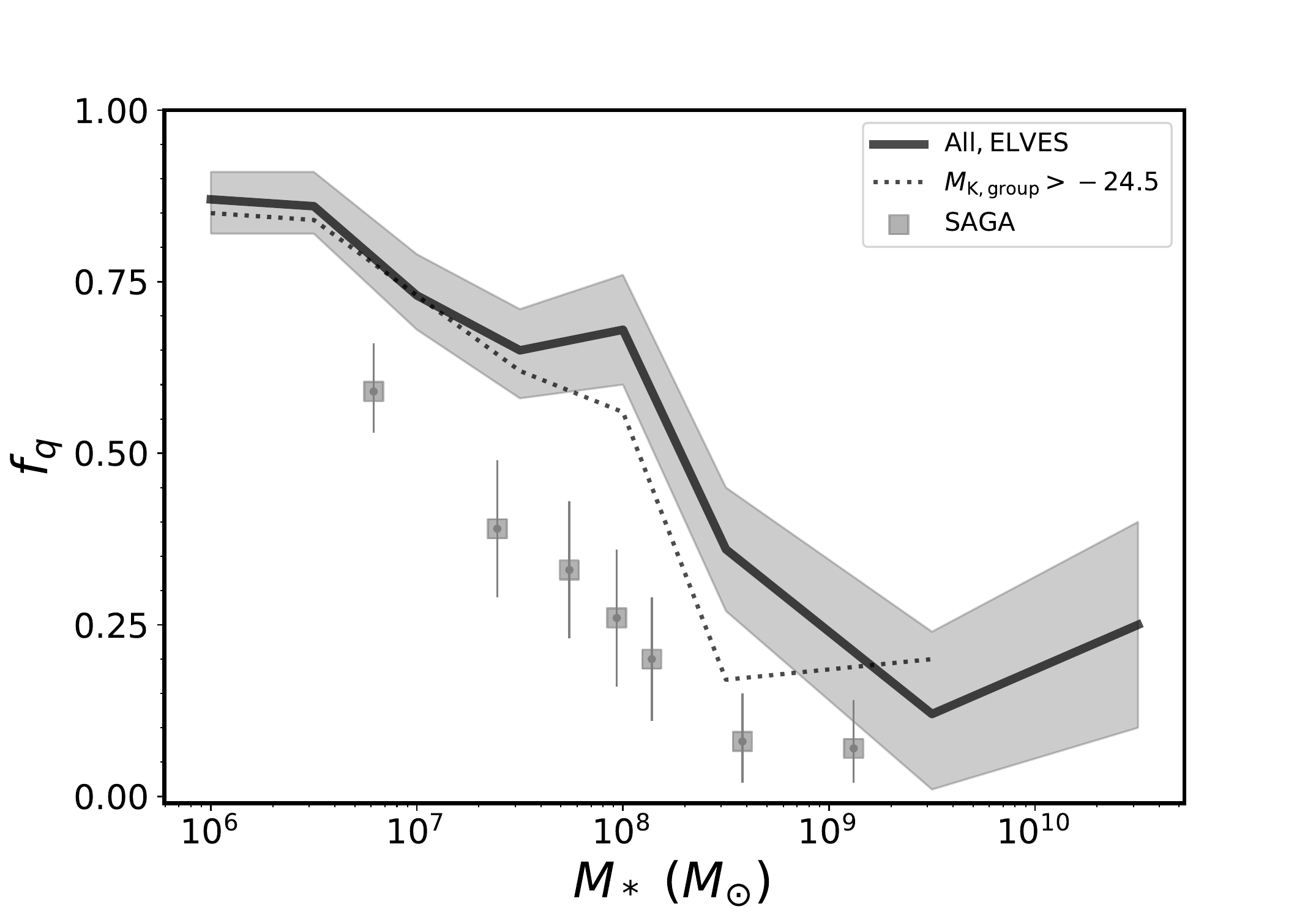}
\includegraphics[width=0.48\textwidth]{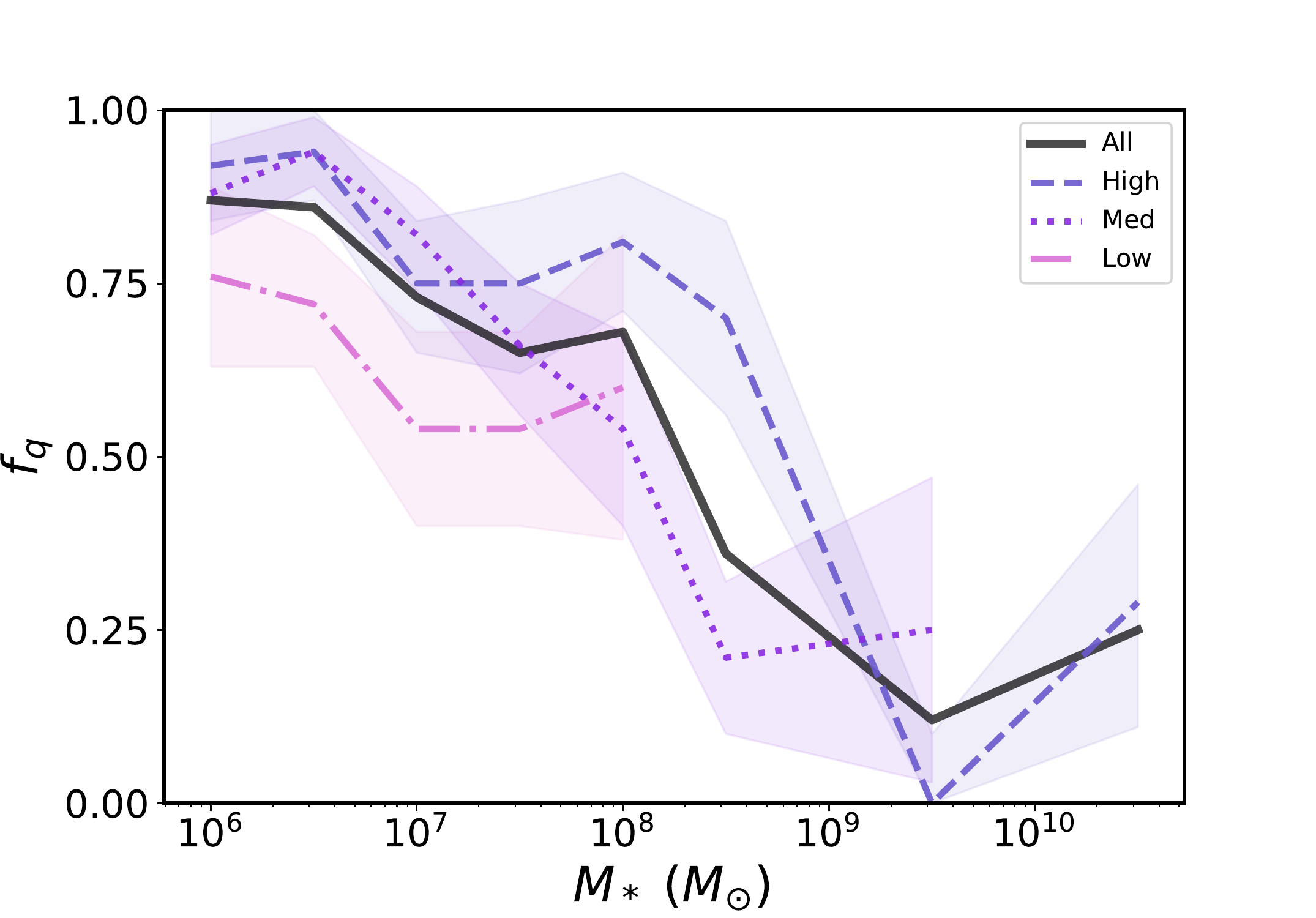}
\caption{{\it Left}: Quenched fraction ($f_q$) with stellar mass for the entire ELVES sample (grey). In dotted, we show the quenched fraction derived for only secure satellites with $P_{\rm sat} > 0.8$. We compare with SAGA (grey squares reproduce Fig.\ 11 from Yao et al.), where star-forming galaxies are identified based on H$\alpha$; see C22 for comparisons between SAGA and ELVES where quenched fraction is determined by color. SAGA is likely missing a small fraction of red satellites. {\it Right}: Quenched fraction as a function of satellite mass in \mkgroup\ bins (see Table \ref{tab:samples}). We see a hint that the highest \mkgroup\ bin quenches the most massive satellites more efficiently for $M_* > 10^8$~\msun.
\label{fig:fqhalobin}}
\end{figure*}

Host halo mass, satellite mass, and potentially radial position within the group, may all play a role in quenching efficiency. Therefore, one way that we can leverage our large statistical sample is to examine trends in satellite \mstar\ and projected host distance $R_p$, within bins of \mkgroup. We show the mass vs.\ radial-distance distributions of early- and late-type satellites for the three \mkgroup\ bins in Figure \ref{fig:sattwod}. Note that in some hosts we have cataloged galaxies beyond $R_p = 300$~kpc, but not in a uniform manner. These satellites are not included in our measured quenched fraction.

Even just from this two-dimensional look at the satellite distributions in \mkgroup\ bins, a number of clear trends emerge. The maximum satellite mass is a function of \mkgroup, which means that we can only measure the quenched fraction in \mstar$> 10^{10}$~\msun\ satellite galaxies in host halos that are more massive than the MW. We can also immediately see that at low satellite stellar mass, and low projected host distance, the satellites are predominantly quenched, while the fraction of star-forming satellites clearly rises at all mass and distance as we move to lower halo-mass hosts. Finally, a little quirk is that in the highest \mkgroup\ bin, we see a dearth of galaxies with $M_* < 10^6$~\msun\ and $r < 100$~kpc. This is likely to be a result of incompleteness at least in part, due to higher contamination from bright galaxies at the group center. On the other hand, there also may be a real dearth of such low-mass satellites towards the group center.

To quantify these apparent trends further, we collapse in radius and examine quenched fractions in satellite stellar mass bins (Figure \ref{fig:fqhalobin}). This figure is very similar to Figure 11 from C22, and again emphasizes that we see a high quenched fraction for \mstar$<10^8$~\msun, and a declining quenched fraction for more massive satellites. These results are highly consistent with (and contain) the MW and M31, as well as other works looking at groups in larger surveys \citep[e.g.,][]{Wheeler:2014,Baxter:2021}. In the ELVES sample, \mstar$\approx 10^8$~\msun\ does look like a transition mass where the quenched fraction falls below 50\%. The results are also consistent with a number of recent simulations \citep{Akins_2021,Samuel_2022,Font_2022}. As a check that we are not biased by the larger groups, we also look at the overall quenched fraction with the four highest-\mkgroup\ hosts (M31, M81, NGC~3379, NGC~3627) excluded. We find a very similar overall trend (shown as dotted line in Figure \ref{fig:fqhalobin}), with only a slightly smaller $f_q$ at satellite \mstar$>10^8$~\msun. 

We also compare with the SAGA survey, as they have reported a significantly lower quenched fraction than in the Milky Way \citep[][]{Geha_2017,Mao_2021}. In this figure, we are plotting the data from Mao et al.\ with a correction for spectroscopic incompleteness applied to the quenched fractions directly. The figure appears slightly different from Figure 11 in Mao et al. We discuss the comparison with SAGA, and the ELVES/SAGA comparison, in the next section.

Looking at the quenched fraction as a function of \mkgroup\ (Figure \ref{fig:fqhalobin}, right) we see that the quenched fractions are similar within $\sim 10-15\%$ across halo mass bins. We do see a hint that higher-mass halos are more efficient at quenching galaxies. Specifically, in the satellite mass bin log~$M_* = 8-9$~\msun, we find $f_q = 70\pm14\%, 20\pm11$ for the High and Medium halos respectively, while at lower stellar masses the quenched fractions are quite consistent across all three halo-mass bins. Larger samples of satellites at the bright end should be relatively straightforward to amass, and would provide additional statistical support to the possible trend we detect here.

\subsection{Comparison with SAGA}
\label{sec:sagacomp}

\begin{figure}
\hskip -5mm
\includegraphics[width=0.5\textwidth]{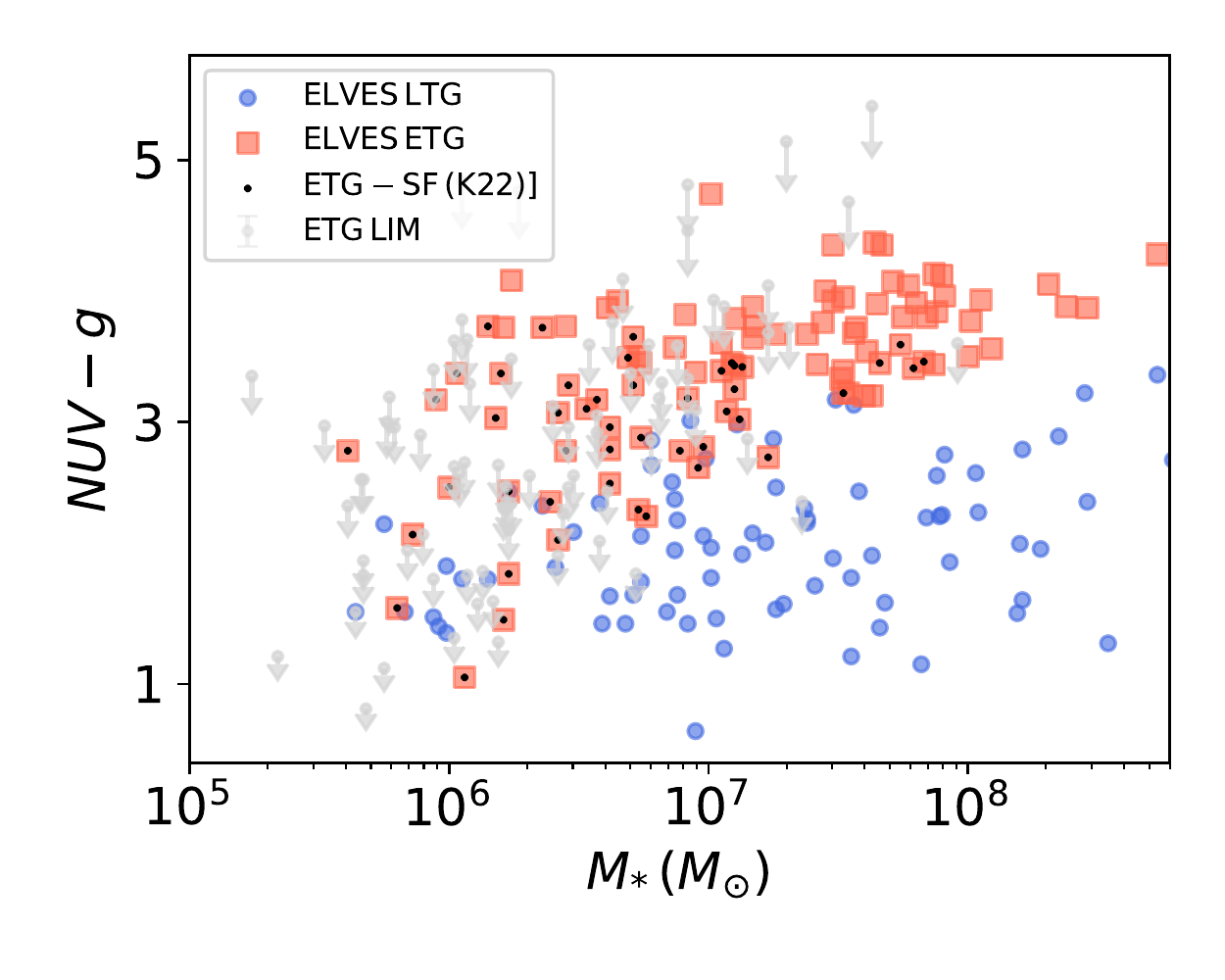}
\caption{The color-mass relation for the ELVES galaxies with \emph{GALEX} coverage (upper limits for ETGs indicated in grey). Note the distinct tilt towards bluer colors for lower-mass galaxies. The galaxies that \citet{Karunakaran:2022elves} would label as star-forming based on their NUV luminosities are indicated with the black dots, showing again that nearly every morphologically early-type satellite with $M_* < 10^7$~\msun\ is considered star-forming by this criterion. As argued in the text, given the star-formation histories that do exist for these galaxies, this metric for star formation seems overly inclusive. 
\label{fig:mainsequence}}
\end{figure}

It is worth considering all the factors that may contribute to the observed differences between SAGA and ELVES; we direct the reader to \S 7.1 of C22 for complementary detailed discussion. Between ELVES and SAGA, the host mass functions are different, in the sense that the halo distribution is skewed to higher mass for ELVES, which may lead to higher quenched fractions at fixed satellite mass (as we argued above in \S \ref{sec:massdist}). There are also differences in mass limit and radial coverage, both of which could push ELVES to higher quenched fraction. Uncertainties in $P_{\rm sat}$ and membership also contribute to uncertainties for both surveys. Finally, we define our quenched samples differently.   

We can mitigate the host differences by including only those hosts that would pass the SAGA cuts, but as shown in Figure \ref{fig:fqhalobin}, the difference in quenched fraction with satellite mass barely change when the most massive hosts are removed. We can control for the differing satellite mass limits by comparing at fixed mass. The radial coverage is close to matching (in projection) as we consider only hosts with coverage at or beyond 200 kpc. We are then left with whether SAGA is missing red galaxies, or ELVES is missing blue galaxies. 

ELVES do see some evidence that SAGA may be missing a small fraction of satellites \citep[see also][]{Font_2022,Karunakaran:2022elves}. C22 carefully matched ELVES and SAGA in satellite luminosity ($M_V \sim -12$~mag), host magnitude distribution, and radial distribution. Even with all of these constraints, we still see very clear differences in the luminosity functions of ELVES and SAGA hosts, with ELVES finding roughly one additional satellite per host \citep[see also][]{Karunakaran:2022elves}. Since, at fixed stellar mass, red galaxies are fainter, overall it seems plausible that SAGA would be preferentially missing some red galaxies. We suspect that a cause of difference between our quenched fractions and those of SAGA are due to these missing galaxies. 

However, there is also the question of whether ELVES is missing star-forming galaxies. SAGA finds a higher number density of blue satellites particularly at the bright end ($M_V \lesssim -15$~mag). As argued by C22, it seems unlikely that ELVES is incomplete in these targets, since satellites this luminous have long been known within the Local Volume; ELVES does not add new sources at these luminosities.

A bigger concern is the way that ELVES determines star formation rates and whether we mis-classify star-forming galaxies as quiescent ones. SAGA has measured H$\alpha$ equivalent widths for all of their satellites. H$\alpha$ obviously probes instantaneous star formation. There is not available H$\alpha$ for all ELVES galaxies, although the morphologically determined active and quiescent galaxies have nearly disjoint distributions of H$\alpha$ (see Appendix). Given the overlap between the H$\alpha$ from morphologically quenched and star forming galaxies, we estimate 10-15\% uncertainty in our quenched fractions if H$\alpha$ were perfectly correlated with star formation.

\citet{Karunakaran:2022elves} take a different approach to identify star-forming galaxies within ELVES. They derive a star-formation rate from the NUV luminosity, and then apply a fixed specific star formation rate cut. They show that the quenched fraction drops dramatically for ELVES under this prescription, particularly at the lowest stellar masses. However, we suspect that the NUV luminosities are a misleading indicator of star-formation rate in these low-mass systems, likely because the conversions are calibrated at higher metallicity. Thus, we contend, the star-formation rates are likely overestimated at low stellar mass. We try to summarize our argument in Figure \ref{fig:mainsequence}, where we show the color-mass relation. The red squares are classified morphologically by C22 as early-type, while the blue circles are late-type. The two sequences converge in color at low-mass. This tilt in the color-mass relation is also shown in C22 and included in our color selection of star-forming galaxies. As argued above, the relative excess of NUV light is likely to be a metallicity effect rather than a sign of excess star formation. Adopting the specific star formation rate threshold from \citet{Karunakaran:2022elves} of sSFR$=10^{-11}$~yr$^{-1}$, we see that we would classify virtually all of the morphologically early-type satellites with $M_* < 10^7$~\msun\ as star-forming. 

There are four ELVES satellites that are morphologically classified as early-type galaxies, have $M_* < 3 \times 10^7$~\msun, have \emph{GALEX} NUV detections, and have color-magnitude--based star-formation histories from \emph{Hubble Space Telescope} \citep{Weisz_2011}. In these four galaxies [BK5N, FM1, f8d1, HS117], also classified by Weisz et al.\ as early-type, the star-formation histories indicate that the galaxies are old. In all cases, the galaxies had formed between 83-97\% of their stars 3 Gyr ago, and 90-97\% of their stars were in place 1 Gyr ago. Although these systems represent a small subset of the early-type galaxies, they do support our suggestion that the NUV emission is not arising from ongoing star formation. We conclude that the $\sim 30\%$ mean quenched fraction suggested by \citet{Karunakaran:2022elves} based on a NUV luminosity is likely to be an overestimate of the star-formation rates in low-mass ELVES galaxies.

\subsection{Quenching and Distance From Host}
\label{sec:fractions_radial}

\begin{figure*}
\includegraphics[width=0.46\textwidth]{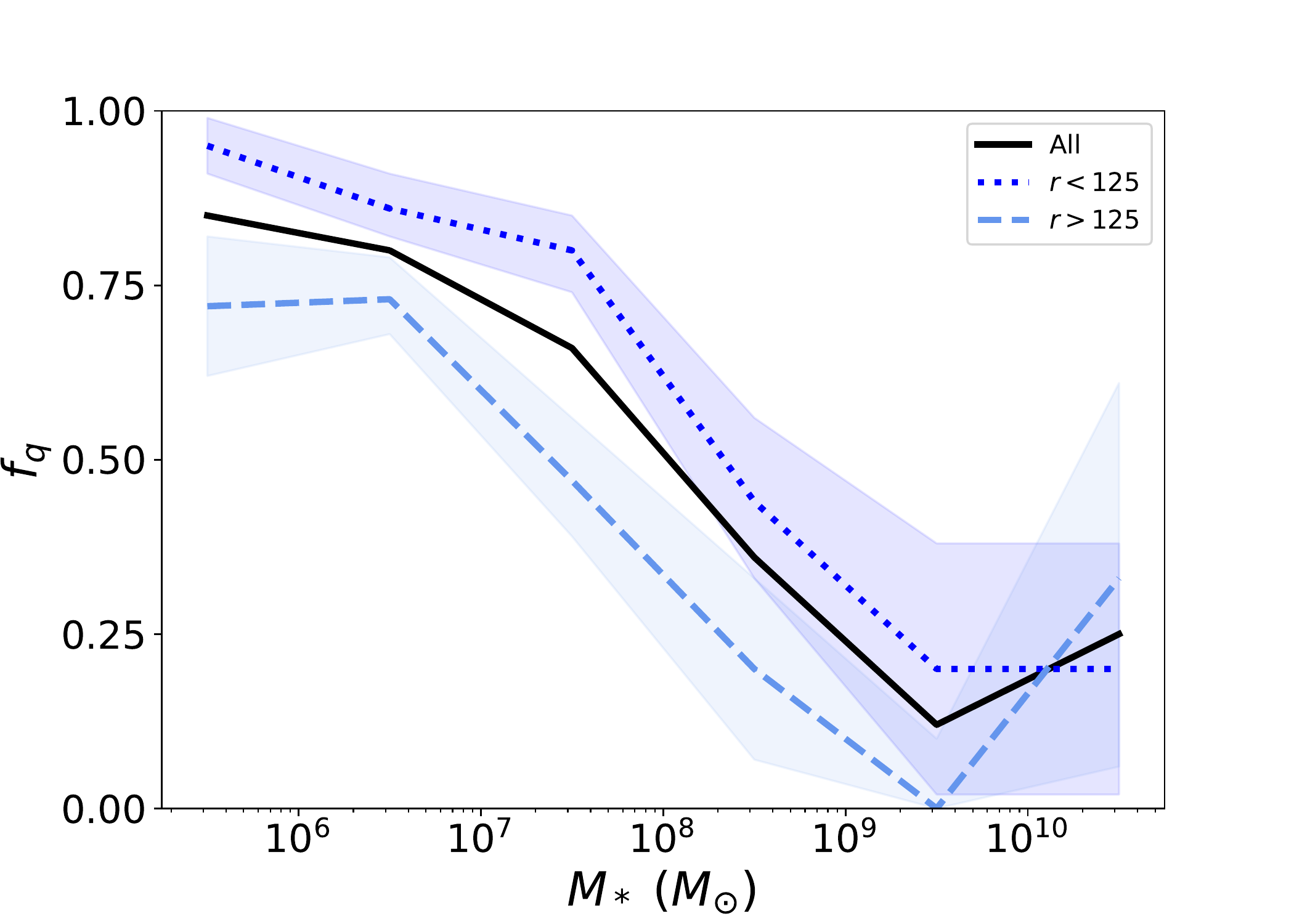}
\includegraphics[width=0.46\textwidth]{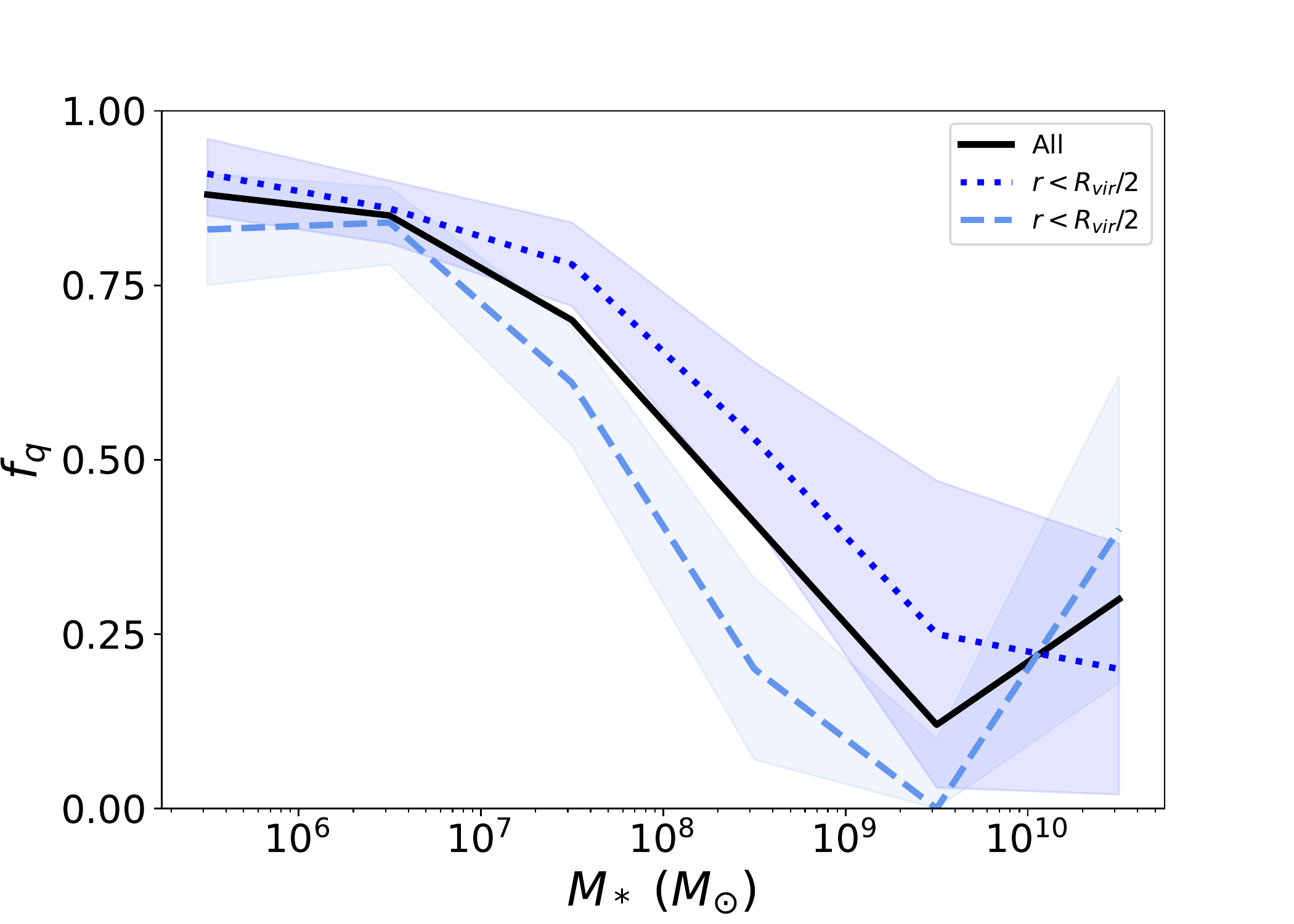}
\caption{{\it Left:} Quenched fraction with stellar mass now also split into two radial bins. We can detect a trend for higher quenching at smaller radius at fixed satellite mass, but only when we investigate the sample as a whole; the statistics are too poor to look at radial dependence in $f_q$ for each \mkgroup\ bin.
{\it Right}: Same as at left, except comparing within and outside of $R_{\rm vir}/2$.
\label{fig:fqmstarrad}}
\end{figure*}

We then examine the quenched fraction with stellar mass, but in two radial bins (Fig.\ \ref{fig:fqmstarrad}). We look both in kpc and $R_{\rm vir}$ bins. First, we divide the sample at a projected distance of 125 kpc, or about one-third of the virial radius. We chose this radius because it roughly divides the samples in half, but our results are unchanged, allbeit with slightly poorer statistics, if we use 150 kpc. We see that at fixed satellite \mstar, the quenched fraction is slightly higher for satellites closer to the host over the full mass range that we probe. However, when we divide by $R_{\rm vir}/2$ (Fig.\ \ref{fig:fqmstarrad}, right), we see that some of the apparent differences are likely driven by mixing in $R_{\rm vir}$ space. Thus we conclude that while there is a hint of increased quenching at smaller radius, more statistics will be needed to get a firm measurement.

It seems intuitive that satellites sitting closer to the host should have earlier infall times and thus a longer time to quench. In detail, \citet{Santistevan:2022} explore the relationship between host distance, infall time, and orbital energy \citep[see also early work by][]{Johnston:2008aa}. They show that while there is a correlation between present-day host distance and infall time, there is significant scatter, and apparently orbits do not necessarily shrink monotonically with time, due to both satellite-satellite interactions and the changing halo potential. Thus, it is not a priori obvious whether we expect to find radial differences in quenching.

\subsection{Quenching and host morphology}

\begin{figure}
\includegraphics[width=0.4\textwidth]{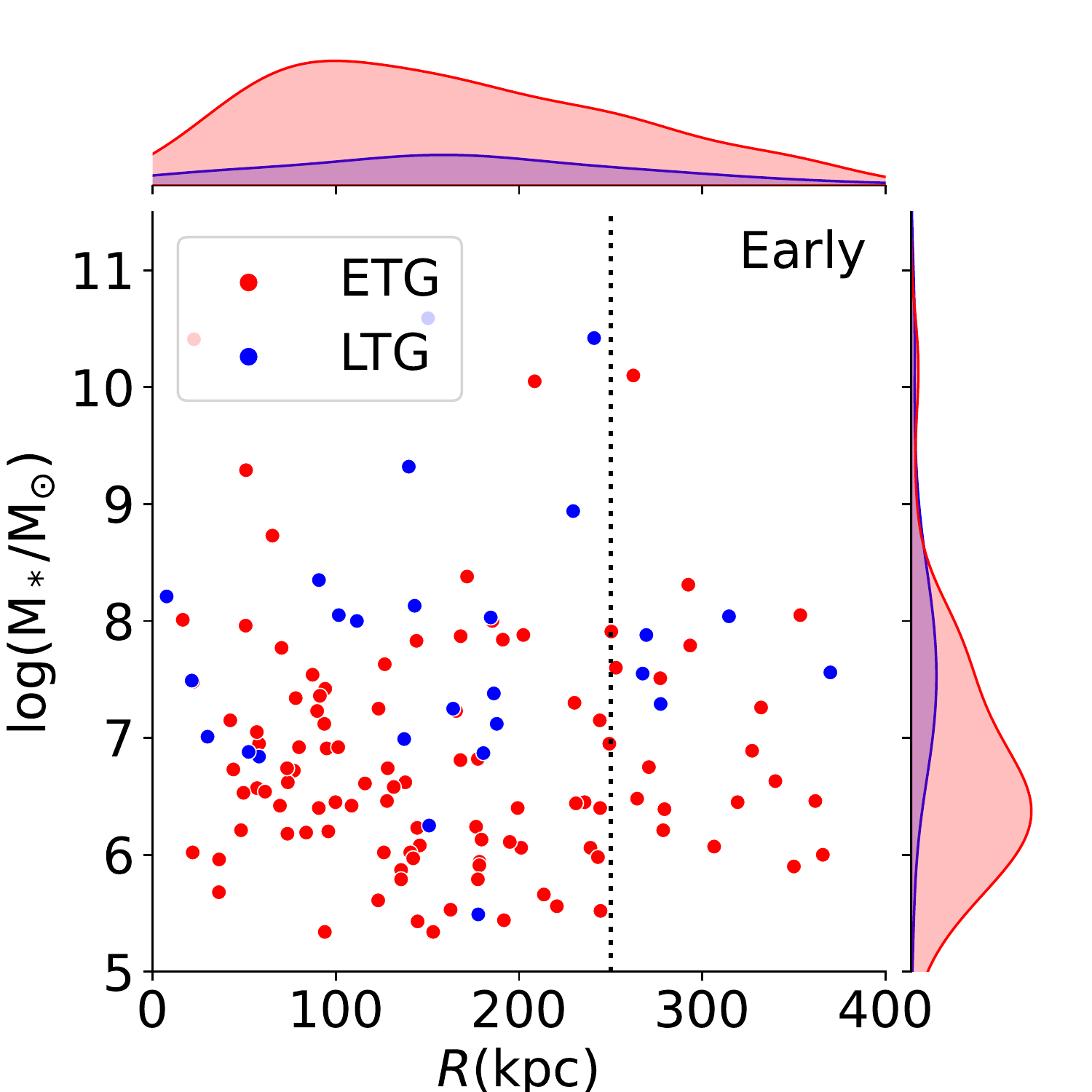}
\includegraphics[width=0.4\textwidth]{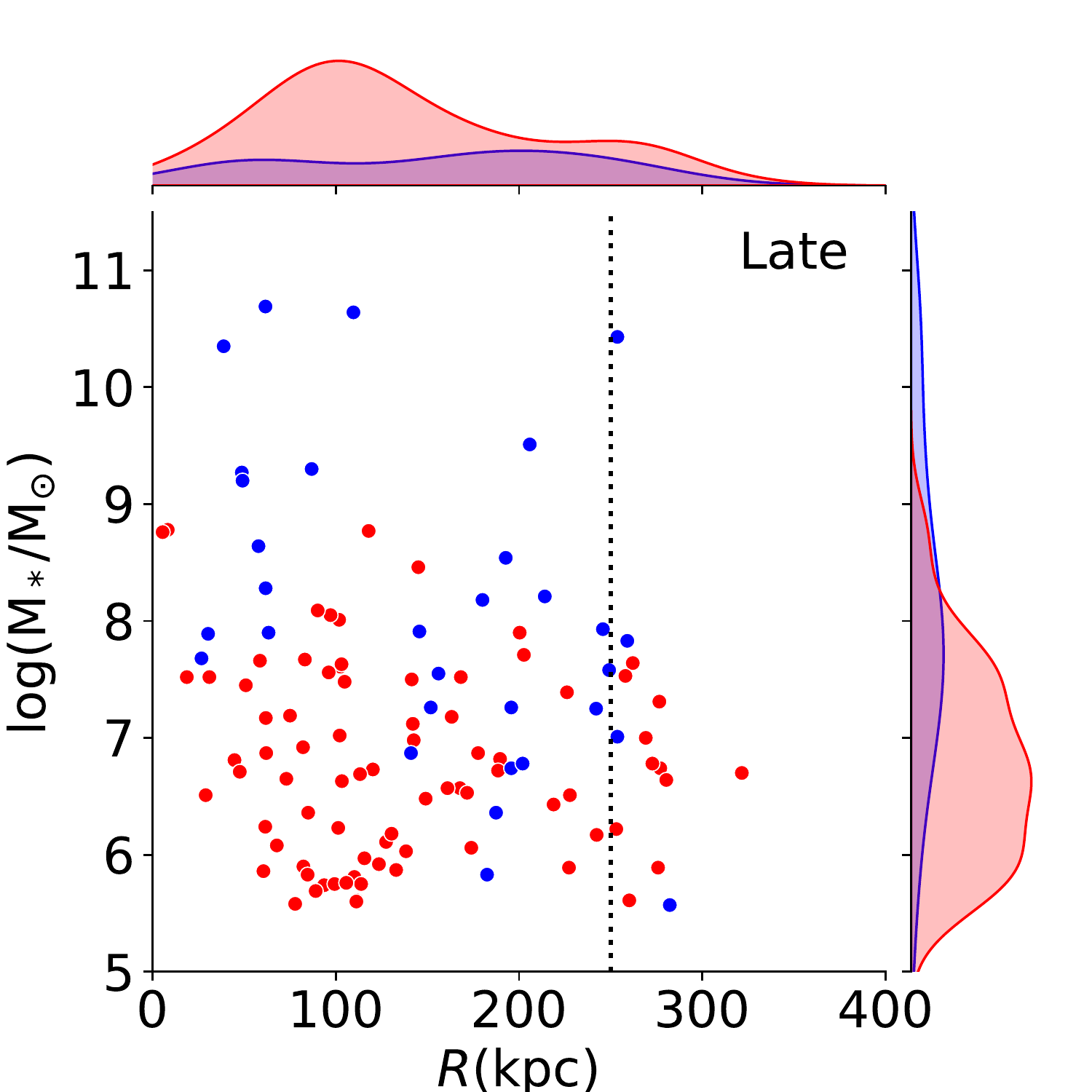}
\caption{ELVES satellites with a high probability of being a satellite $P_{\rm sat}>0.5$ around the hosts with \mkgroup$<-24$ and divided into early and late-type hosts. Only galaxies within a projected distance of 250 kpc are included in the quenched fractions. We distinguish between satellites that are ETG (red) and LTG (blue). Any differences between the two groups based on morphology are subtle. 
\label{fig:sattwod-earlylate}}
\end{figure}

Thus far, we have aggregated satellites of early-type and late-type hosts. We have marginally enough data to address whether there are differences in satellite quenching between elliptical and spiral hosts. We attempt to draw from comparable halo mass distributions by taking the five early-type hosts (NGC~1023, NGC~1291, NGC~3115, NGC~3379, Centaurus A) and then drawing the seven late-type hosts down to a matching \mkgroup$<-24$~mag. Of course, we cannot know for sure that we have similar halo mass distributions across the two samples given their morphological differences \citep[][]{Kauffmann:2013}, but it is interesting to investigate whether we can detect differences across the sample based on host morphology. We show the two-dimensional mass and radial distributions of the satellites from the early and late-type hosts separately in Figure \ref{fig:sattwod-earlylate}. There are more satellites overall around the early-type hosts, and they extend to larger radius. However, for uniformity, we restrict our attention to satellites within 200 kpc. 

\begin{figure}
\includegraphics[width=0.48\textwidth]{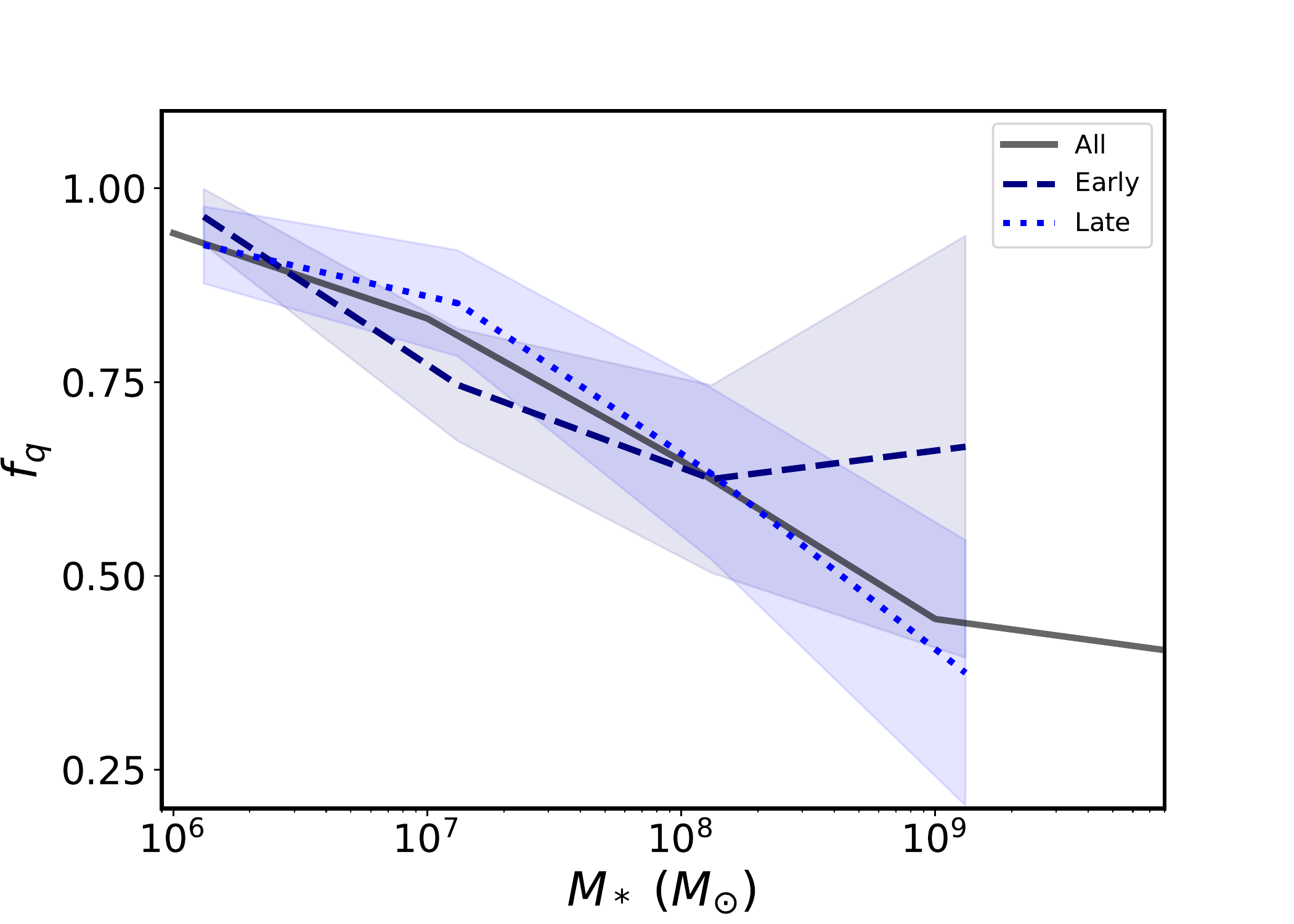}
\caption{ELVES satellite galaxies in around early vs late-type hosts, including all ELVES hosts with $M_{\rm K, group} < -24$~mag. Five early-type hosts and seven late-type hosts are included in this comparison (including the MW and M31 in the late-type hosts). 
\label{fig:fqmorphology}}
\end{figure}

In Figure \ref{fig:fqmorphology}, we compare the quenched fractions between the early- and late-type hosts. We do not see a measurable difference between the two. At a given mass, the quenched fractions of the early and late-type hosts in our sample are similar at the $\sim 10\%$ level. We will compare with prior work on ``galaxy conformity'' in \S \ref{sec:discussion}.

\section{Quenching Times}
\label{sec:satgen}

By comparing measured quenched fractions with infall time distributions from cosmological models, we can infer an average time to quench \citep[e.g.,][]{Wetzel_2015,Fillingham_2016}. A simple abundance-matching scheme links our data to the satellite distributions predicted by the semi-analytic model \satgen\ \citep{Jiang_2021_satgen} to yield infall time distributions as a function of satellite mass. We describe the details below.

\begin{figure*}
\includegraphics[width=0.45\textwidth]{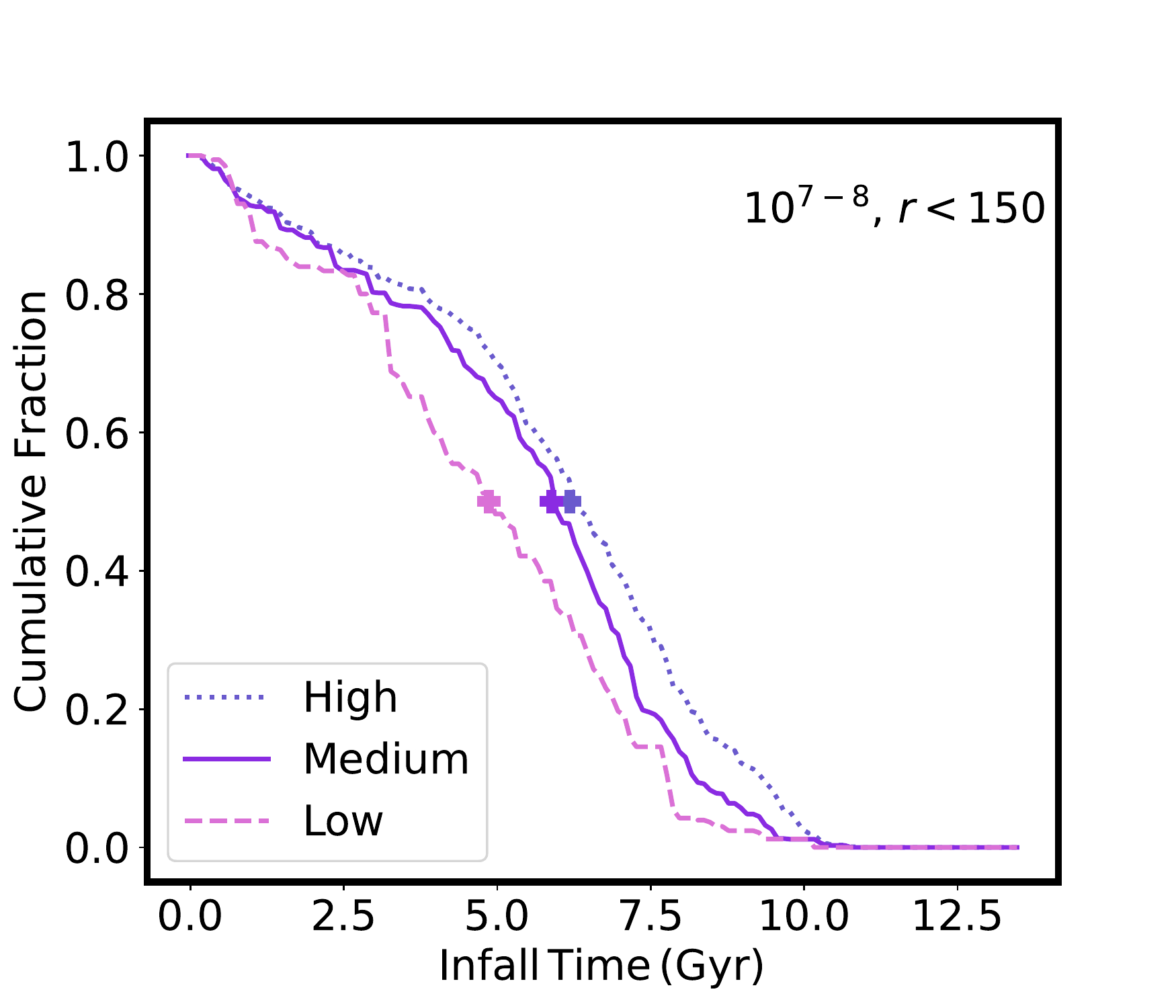}
\includegraphics[width=0.45\textwidth]{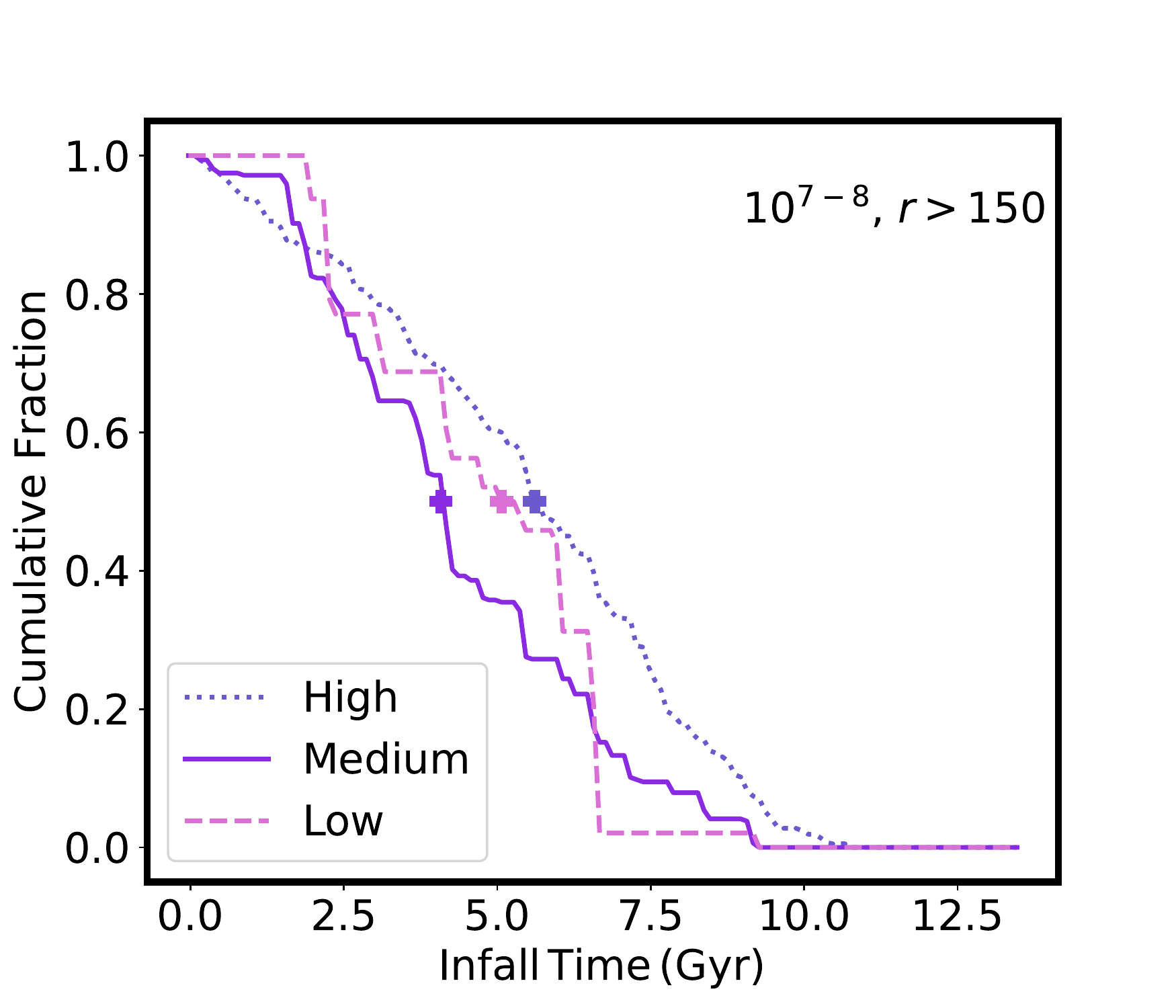}
\caption{Infall time distributions for subhalos with the same distribution of $M_*= 10^{7-8}$~\msun, into the three \mkgroup\ bins. These infall time distributions are combined with the observed quenched fractions to calculate a mean quenching time.
\label{fig:tinfall}}
\end{figure*}

\subsection{\satgen\ Models}

\citet{Jiang_2021_satgen} present a semi-analytic model of sub-halo evolution within a MW-mass halo. The code uses very high-resolution zoom-in hydrodynamical simulations to build analytic prescriptions for the effects of dynamical friction and tidal stripping on satellites, but runs rapidly, such that it is possible to build up statistical samples over different scenarios in reasonable computational time. Because it encodes the results of very high-resolution hydrodynamical runs, \satgen\ is able to reproduce the radial distribution of observed satellite systems, whereas cosmological simulations do not \citep[e.g.,][]{Carlsten:2020radial,Kravtsov:2022}. At the same time, it is possible to build up large samples for statistical comparisons over multiple host halos without prohibitive computational expense, so that our \satgen\ simulations are ``tuned'' to the halo mass distribution of our host sample in a way that has not been possible with prior cosmological zoom simulations.

The particular SatGen runs adopted in this work emulates a ``strong-feedback'' hydrodynamical simulation. In particular, the impact of stellar feedback on the halo structure of the satellites is based on that in the Numerical Investigation of Hundred Astrophysical Objects (NIHAO) simulations \citep{Tollet:2016,Freundlich:2020},
and is very similar to that in the  Feedback In Realistic Environments (FIRE) simulations as well \citep{Lazar:2020}. The bursty star formation and feedback in the FIRE models make the massive dwarfs puffy and cored, and thus more susceptible to tidal disruption. SatGen can alternatively emulate simulations of weak/smooth feedback, which do not produce cored dwarfs. These satellites are more resistant to tidal effects, but we do not test the smooth feedback models in this work. The strong-feedback emulator as used here makes the satellite mass function $\sim25$\% lower than with smooth feedback, considering satellites more massive than $M_{\star} > 10^5$~\msun\ \citep[][their Fig.\ 4]{Jiang_2021_satgen}. It causes more disruption of small-pericenter satellites and makes the overall radial distribution of satellites less concentrated. However, the effect is subdominant compared to the scatter in the spatial distribution of satellites caused by the dramatic halo-to-halo variance due to differing merger histories, which is captured by our SatGen runs. 

We have recently completed a study of the stellar-to-halo-mass relation (SHMR) through abundance matching between ELVES and \satgen\ \citep{Danieli:2022}. Danieli et al.\ compute a suite of \satgen\ runs for each ELVES host, picking host halo masses from a normal distribution around the mean stellar-to-halo-mass relation of \citet{Rodriguez-Puebla:2017} with a scatter of 0.15 dex. They populate these halos with galaxies according to a satellite-subhalo connection model, utilizing a relation of the form $M_{\star} \propto M^\alpha_\mathrm{peak}$, with a mass-dependent scatter \citep[motivated by the findings of][]{Nadler:2019}. Finally, Danieli et al.\ carefully forward-model all ELVES-related selection effects, including mass and surface-brightness incompleteness, mass bias, distance determination ambiguity and inhomogeneous radial coverage per host. They are able to investigate the host-to-host scatter and constrain the relation with far better statistics (in our satellite mass range) than papers based on the Milky Way alone \citep[e.g.,][]{Garrison-Kimmel:2017,Nadler:2019}. In this work, we use the resulting SHMR from Danieli et al.

We draw 50 simulated halos to match each ELVES-host from the \satgen\ suite described above. Then, to populate these halos, for each draw we rank the simulated satellites by $M_\mathrm{peak}$. We take $M_{\rm peak}$ because tidal stripping will change the halo mass as the satellite orbits in its bigger host-halo. We then select the most massive $N$ subhalos, where $N$ is the number of observed satellites in that ELVES host, to derive a statistical rendition of each ELVES host, that includes predicted masses, radii, and (crucially, for this work) infall times for its population of satellites\footnote{We alternatively took the forward-modeled satellite samples, including observational incompleteness, directly from Danieli et al., again using 50 realizations per host. We do not find different results deriving the infall time distributions in that way.}. Finally, each satellite is assigned a stellar mass from the Danieli et al.\ stellar-to-halo mass relation.

The distribution of infall times from \satgen\ for the satellite stellar mass range of \mstar$= 10^{7-8}$~\msun\ is shown in Figure \ref{fig:tinfall} as a function of present-day radius within the simulated halo. Infall time in \satgen\ is defined at the moment a satellite first crosses the host virial radius, which is the same moment that the halo mass $M_{\rm peak}$ is defined. We should note here that we do not have full observational coverage to the virial radius for all ELVES hosts, although $R_{\rm vir} \sim 250-300$~kpc is probably very close to the virial radius for the MW-mass hosts. When making a radial cut in projection, some satellites will have larger true three-dimensional separations and artificially join our sample in projection; thus, we will naturally have some interlopers. 

Based on \satgen, the typical satellite in this mass range has spent roughly 5-7 Gyr within its current host halo. The distribution of infall times with host mass, satellite mass, and host separation depends on at least three factors. First, more massive halos, in general, assemble later. Second, the accretion history of satellites will be self-similar when normalized by host mass, such that satellites of a given mass are accreted relatively later (earlier) into a less (more) massive host halo. Therefore, at a fixed satellite mass, the typical satellite will have orbited longer in a more massive host. And third, the radial trends should all be compared normalized to the virial radius, which is not strictly possible for our sample. This is why we rely on \satgen\ to deliver infall time distributions accounting for host halo mass, satellite mass, and radial distribution.

For our default model, we assume that every galaxy looks quenched after exactly the same amount of time ($t_q$) within the host halo. In simulation papers, this is often referred to as the quenching delay time. If $t_q$ is short, as argued for the MW \citep[ $\sim$2~Gyr, e.g.,][]{Fillingham_2015}, the quenching is likely due to ram-pressure stripping that can act on those shorter time scales. Once the $t_q$ grows long, the galaxy may quench primarily via starvation, i.e.\ from running out of access to additional fuel \citep{Fillingham_2016}. For reference, in the case of the MW, there appears to be a discontinuity in the quenched fraction with satellite stellar mass, whereby satellites with \mstar$<10^8$~\msun\ are nearly all quenched, implying that the quenching mechanisms act rapidly in those cases, while galaxies at higher mass are mostly unquenched, implying that those systems are able to retain their gas reservoirs \citep[e.g.,][]{Grcevich_2009,Spekkens_2014,Fillingham_2015,Putman_2021}. With the sample of 30 MW-like hosts in ELVES, we can investigate whether the apparent sharp mass division in the dominance of these two physical processes is a feature of the MW-mass hosts or is a unique element of the MW's accretion history.

\begin{figure*}
\includegraphics[width=0.45\textwidth]{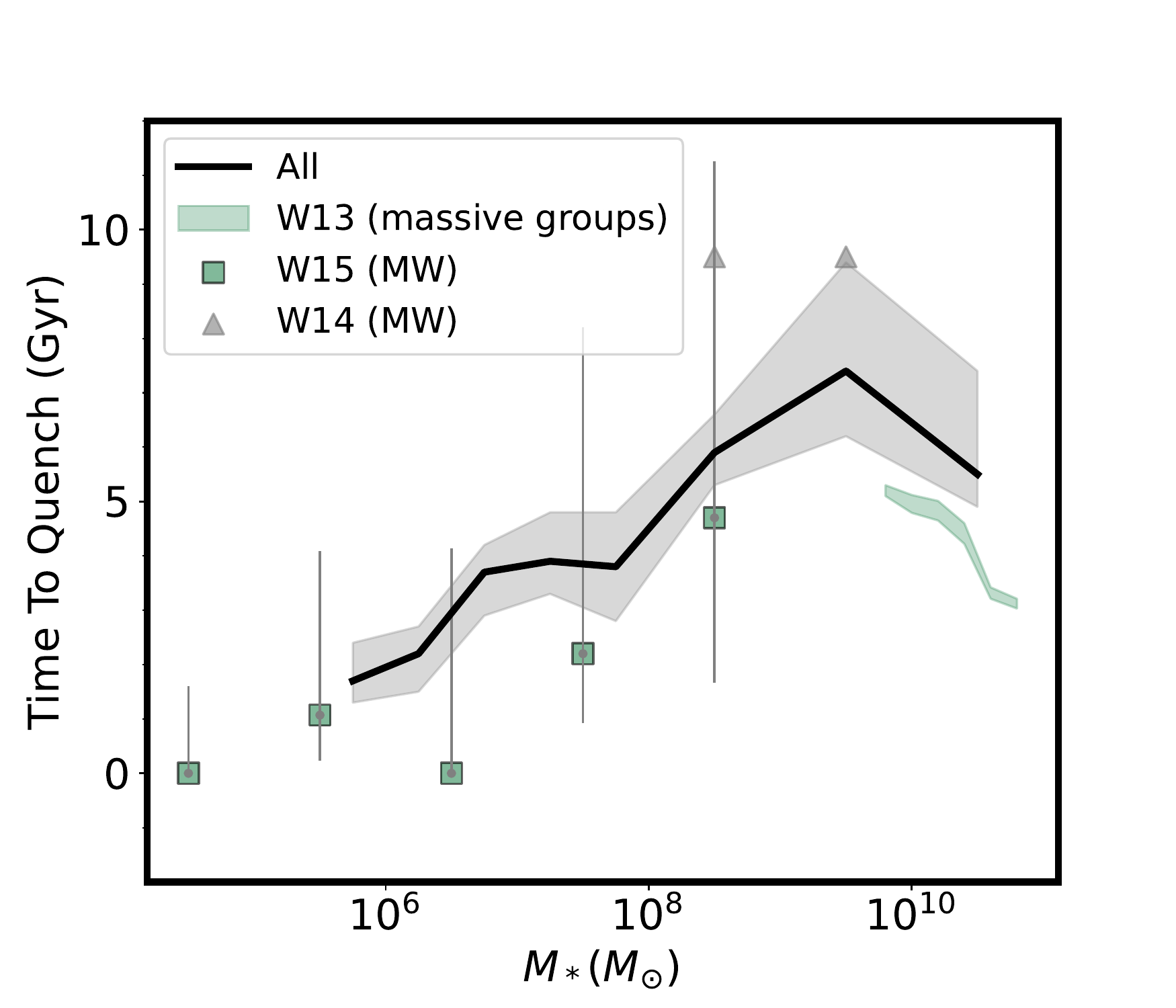}
\includegraphics[width=0.45\textwidth]{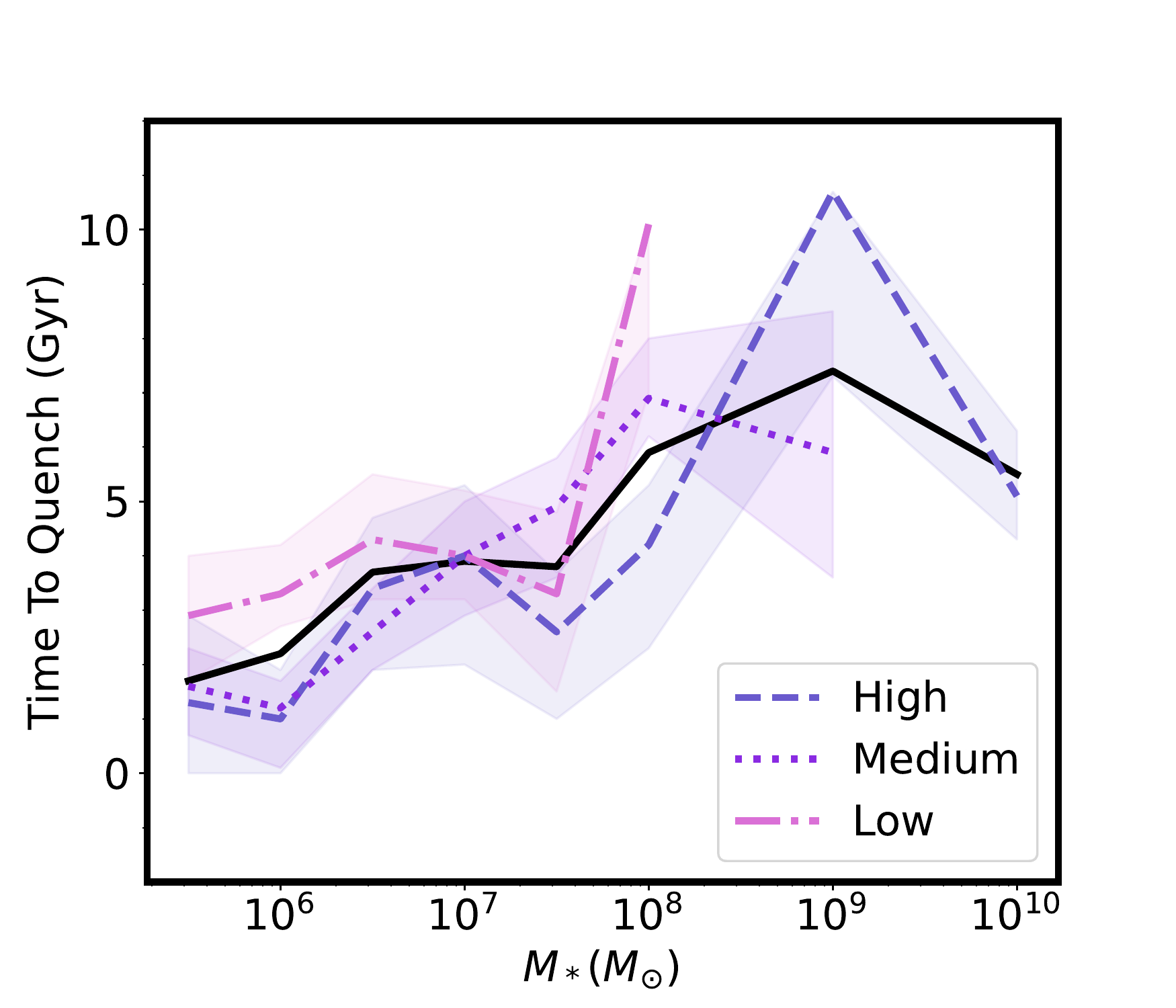}
\caption{Inferred quenching time in satellite stellar mass bins for the full ELVES sample (left) and in bins of \mkgroup\ (right). The time is inferred by taking the infall time distributions for the \satgen satellites in that bin, and then identifying the $t_q$ value that reproduces the observed quenched fraction. We compare with a similar inference for MW satellites from \citet[][(W15)]{Wetzel_2015} and \citet[][(W14)]{Wheeler:2014}, as well as measurements for more massive groups from \citet[][(W13)]{Wetzel_2013}.
\label{fig:quench}}
\end{figure*}

\subsection{Time to quench}
\label{sec:derivetq}

With an infall time distribution that matches our observed satellite mass-function in hand, we derive a quenching timescale $t_q$ in bins of satellite mass. We assume that every satellite in a given mass bin with an infall time longer than a characteristic $t_q$ is quenched, in order to match the observed quenched fraction per bin. This quenching time, in bins of satellite stellar mass, is shown on the left-hand side of Figure \ref{fig:quench}. We have checked that if we do the matching within a fixed $R_{\rm vir}$-defined radius instead, we recover very similar results.

We see that our inferred quenching time is consistent with the predictions of rapid quenching under ram-pressure stripping for satellites with \mstar~$\approx 10^6$~\msun, but then rises steadily, reaching $t_q\sim 4-5$~Gyr for satellite \mstar~$\approx 10^7$~\msun. The quenching time gets even longer for galaxies with \mstar~$>10^8$~\msun. While in general our results are consistent with those inferred for the MW (\S \ref{sec:discussion}), with the ELVES composite sample of MW-like hosts, a much more gradual increase in average quenching time with satellite stellar mass is revealed than has been inferred for the MW alone.

Given the hints of empirical differences between quenched fractions in \mkgroup\ bins (see Figure \ref{fig:fqhalobin}), we also derive $t_q$ as a function of satellite stellar mass separately for each of the \mkgroup\ bins; the results are shown on the right-hand side of Figure \ref{fig:quench}. To be clear, we are not directly modeling the stripping processes, we are simply calculating the average time since infall required to reproduce the observed quenched fractions given the infall time distributions. The quenching times are quite similar for the different \mkgroup\ bins. 

\begin{figure}
\includegraphics[width=0.45\textwidth]{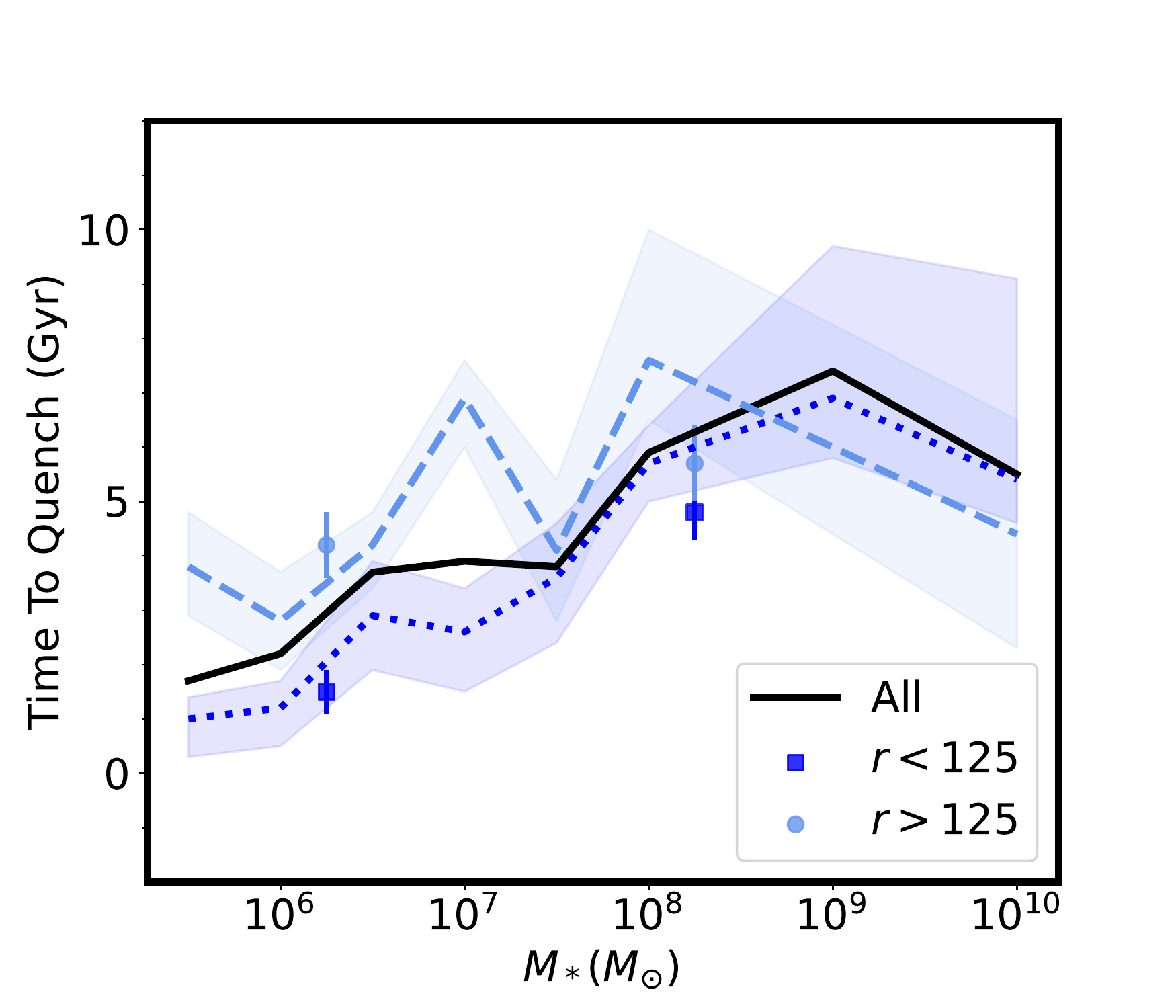}
\caption{Inferred quenching time in satellite stellar mass bins for the full ELVES sample divided by projected host distance. We use the full sample to boost signal-to-noise, but see the same trend of shorter/longer quenching times at smaller/larger projected distance in each \mkgroup\ bin. 
\label{fig:radialquench}}
\end{figure}

We also derive quenching times in the two radial bins from Fig. \ref{fig:fqmstarrad} above in Figure \ref{fig:radialquench}. We use the $z-$axis (line-of-sight) present-day radius within \satgen\ to mimic observed projection effects within the simulation. Investigating the inferred $t_q$ for radial bins directly could allow us to separate two effects -- higher circumgalactic medium density and larger number of pericenter passages closer to the host -- that may contribute to increased quenching for galaxies at smaller projected distance. We see tentative evidence for longer quenching times at larger radius, but this difference is only significant when we average over a large mass range of \mstar$<10^{7.5}$~\msun. In order to explore the uncertainties more completely, we create 5000 jack-knife samples in each bin by selecting from the ELVES sample in that mass bin with replacement. Higher-mass satellites seem to have comparable quenching times regardless of radius, but we see some evidence for longer quenching times at smaller radial distance for the lower-mass galaxies (points in Figure \ref{fig:radialquench}). We caution that this effect may be due to subtle changes in the mass function between the two samples. Thus we consider this result tentative with the current sample. If it is true that low-mass satellites quench more quickly when closer to the host, it would suggest that the higher circumgalactic medium density seen by closer-in satellites shortens their quenching time \citep[see also][]{Simpson_2018}.


\subsection{Caveats}

We want to highlight some important caveats here, since our quenching ``model'' is very simplified. First, not all satellites will quench in the same amount of time in the halo \citep[e.g.,][]{Weisz_2015,Akins_2021,Fillingham:2019}. Thus, when we measure a quenching time longer than the typical 2~Gyr, we are measuring the average over what is likely a wide distribution of quenching times in the real galaxies. We also are measuring quenched fractions in projected radii, which artificially boosts the number of galaxies at larger distance that get projected into the sample. Since the quenched fraction drops with host separation, projection likely lowers the quenched fractions.  Although we try to mimic the projection effects with \satgen, we do not include interlopers that are outside the virial radius but appear to be at the host distance given our SBF precision \citep[see discussion in ][]{Carlsten:2020survey,Carlsten:2022survey}. 

Another complication, secondly, is that we ignore the life of the satellite halo before falling into the host. Some of these satellites may have been quenched in their prior environments, maybe as much as 30-40\% \citep[e.g.,][]{Kravtsov:2004,Wetzel_2015,Diemer:2021,Santistevan:2022}. Thus, it is quite possible that our derived times are too long, if some satellites come in pre-processed (see \S \ref{sec:discussion}), even in dwarf groups \citep[e.g.,][]{Stierwalt_2015,Stierwalt_2017}.

Thirdly, the detailed accretion histories for individual halos probably matter. While \satgen\ is capable of capturing the full cosmological variance in accretion histories, we currently have no way to tune the \satgen\ models per galaxy to account for its larger scale environment or accretion history. Infall time distributions may be different for an M81 group than NGC 3379, for instance. Also, as far as our quenching assumptions, it is clear from \citet{Putman_2021} that for the MW, specifically, the fact that we are part of a group with M31 makes a difference in how efficiently we quench our satellites. Putman et al.\ argue that this is due to a circumgroup medium that starts the ram-pressure stripping process earlier than were it a single-galaxy halo. In our current methodology, we are not treating systems differently if they have multiple massive group members.

We will revisit some of these issues in \S \ref{sec:discussion}.

\section{Discussion}
\label{sec:discussion}

We have investigated the stellar mass and radial dependence of quenching for the satellites cataloged by the ELVES survey. Overall, the quenched fraction is a strong function of stellar mass, with very high quenched fractions for $M_* < 10^8$~\msun. There may be a residual trend at fixed $M_*$ that galaxies are more quenched at lower projected host distance. We also see moderate evidence that the highest \mkgroup\ halos are more efficient at quenching the high-mass satellites with $M_* > 10^7$~\msun.

\subsection{Comparison With Prior Observations}

We have already established that our inferred quenching times are very consistent with those published for the MW and M31 \citep[e.g.,][]{Fillingham_2015,Wetzel_2015}. We also compare with a few observations based on groups in the SDSS. \citet{Wheeler:2014} present upper limits on quenching time for satellites with \mstar$>10^8$~\msun\ in MW-mass groups. Their upper limits on $t_q$ are somewhat higher than our inferred $t_q$ in their lower-mass bin. However, the quenched fractions presented in \citet[][]{Wheeler:2014}, which are based on \citet{Geha_2012}, are $\sim 25-30\%$, for a satellite mass range of $10^8-10^{9.5}$~\msun, which is quite consistent with our measurement (Fig.\ \ref{fig:fqhalobin}), suggesting that the difference is due to different modeling assumptions.

In Figure \ref{fig:quench}, we also show $t_q$ measured around more massive groups $M_h \sim 10^{12}-10^{13}$~\msun\ with SDSS by \citet{Wetzel_2013}. While we do not have enough high-mass groups to robustly compare, there is an intriguing hint that the quenching time is turning over for our highest-mass satellites as in the \citeauthor{Wetzel_2013} results. Our highest-mass bin is entirely dominated by satellites in small groups, like M81, that likely fall in the same halo mass bin, so comparison seems fair. Our inferred quenching times are longer, but we are consistent with that work at the $2 \sigma$ level.

We now return to the interesting result that galaxies of early and late-type morphology have similar quenched fractions in our ELVES sample. Prior work has uncovered differences in the late-type fraction of more massive satellites, with the sense that if the central galaxy is of late-type, the satellites are more likely to be late-type as well. This effect is known as ``galaxy conformity'' \citep[e.g.,][]{Weinmann:2006}. It may be a reflection that at fixed stellar mass, early-type hosts occupy older and more massive dark matter halos \citep{Wang:2012}.

We do not measure a significant galaxy conformity effect in the ELVES sample. We highlight three possible explanations for the lack of detection. One is that we do not have the statistics of these earlier SDSS-based studies, and thus simply cannot measure the galactic conformity signal. The second family of possibilities is with our specific hosts. It is possible that conformity is not strong enough to measure in hosts of this mass, or that we have not put together a halo-matched sample of quenched and star-forming hosts. The final explanation is that the ram-pressure stripping that likely drives quenching in low-mass galaxies is not the only (or dominant) cause of morphological transformation in more massive satellites, so that the physics causing conformity simply does not apply in our satellite mass range. Indirect evidence for this possibility is found in \citet{Carlsten:2021structure}, who also find that red and blue satellites in ELVES obey the same mass-size relation, again suggesting that the morphological transformations that accompany star formation quenching at high mass are not identical to satellites with \mstar$<10^9$~\msun.

\subsection{Comparison with simulations}

\begin{figure}
\hspace{0mm}
\includegraphics[width=0.45\textwidth]{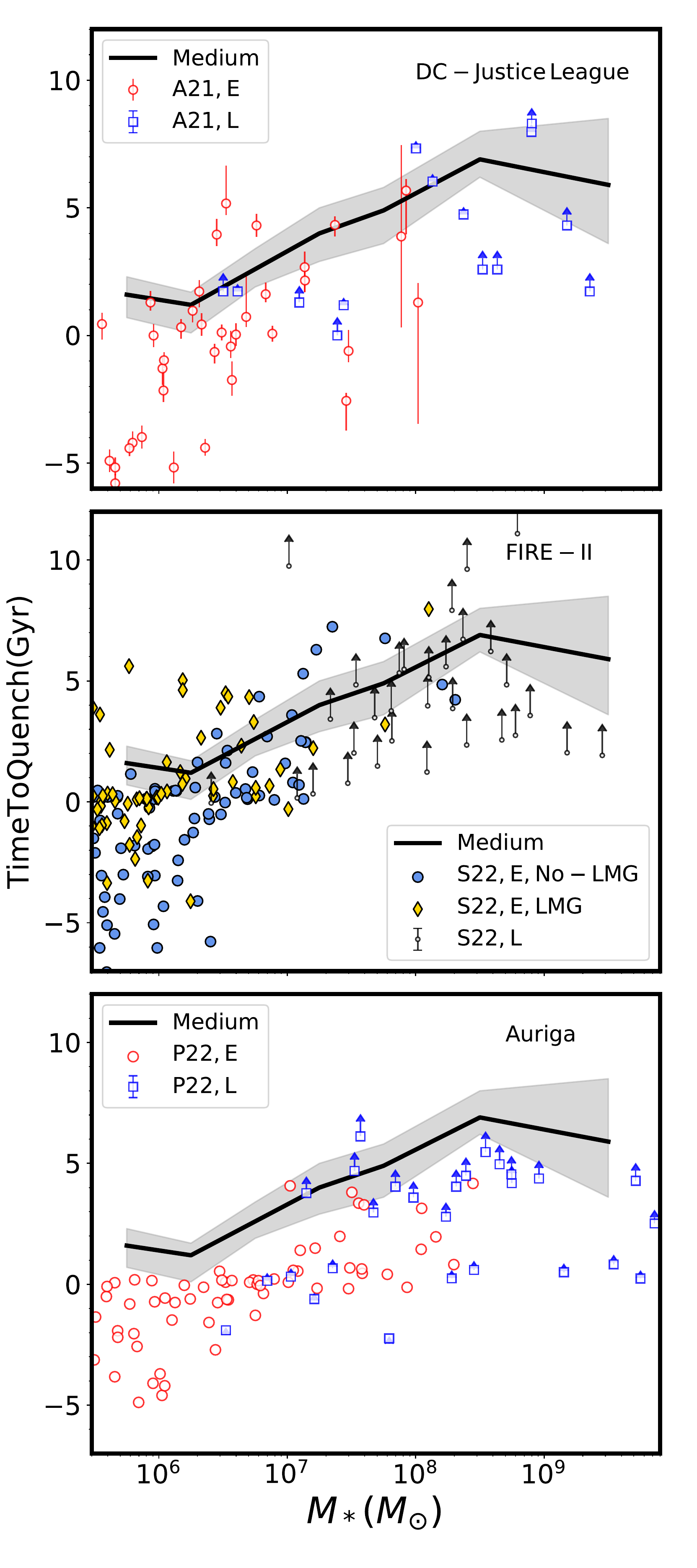}
\vspace{-5mm}
\caption{Time to quench (from hydrodynamical simulations), the difference between infall time and quenching time, often called quenching delay time. We compare delay times as measured from the DC-Justice League simulations \citep[][top]{Akins_2021}, the FIRE-II simulations \citep[][middle]{Samuel_2022}, and the Auriga simulations \citep[][]{Simpson_2018,Pan:2022}. We compare with the inferred quenching times for the ``Medium'' ELVES sample, which is closest to the halo mass range of the simulations. In the case of the Samuel work, they identify two families of satellites, those which were pre-processed in a low-mass group (LMG) and those which were isolated until accretion onto the MW (No-LMG). The LMG satellites are quite often quenched before infall. For the Auriga simulations, we show only the Level 3 (higher resolution) simulations for clarity, and we note that the quenching delay times are slightly shorter in that case because they report $t_{\rm infall} - \tau_{90}$, the latter being the time when 90\% of the stars are formed.
\label{fig:quenchhydro}}
\end{figure}

A number of hydrodynamical zoom simulations have looked at the fraction of quiescent galaxies as a function of satellite mass \citep{Simpson_2018,Buck:2019,Akins_2021,Karunakaran_2021,Font_2022,Samuel_2022}. Overall, the dependence of quenched fraction on stellar mass looks very similar across all the simulations, and agrees with our measurements for the Medium host mass bin \citep[for a summary, see Fig. 13 in ][]{Samuel_2022}. All find very consistent quenched fractions that are near unity for satellites with \mstar$<10^7$~\msun, drop to 50\% between \mstar = $10^{7.5}-10^8$~\msun\ and are near zero for \mstar$> 10^9$~\msun. 

We can also compare our inferred $t_q$ with those from the hydrodynamical simulations. In the simulations, there is perfect knowledge about both infall time and quenching time, and $t_q$ (or quenching delay time) is measured as the difference between the time of infall and the time that star formation ceases. A negative $t_q$ means that the satellite was quenched when it fell into the current host.

In Figure \ref{fig:quenchhydro} (top), we compare with \citet{Akins_2021}, based on the DC-Justice League high-resolution suite of four MW-mass halos \citep{Munshi:2021}. In the middle, we compare with \citet{Samuel_2022} based on models of 14 isolated MW-like or MW/M31-like pairs using the Feedback in Realistic Environments \citep[FIRE][]{Wetzel:2016,Garrison-Kimmel:2019,Hopkins:2018} simulations. At bottom, we compare with the Auriga simulations, their higher resolution (Level 3) runs are shown \citep{Simpson_2018,Pan:2022}. For quenched early-type satellites, we show the time between infall and quenching, while for star-forming late-type satellites, their time since infall is shown as the lower-limit to their quenching time. The one exception is the Pan et al.\ objects for which we show the difference between $\tau_{90}$ (the time for 90\% of the current mass to form) and infall time. 

In the case of Akins et al.\ and Pan et al., the infall time is computed as the first time the satellite passes within the virial radius of the current MW-mass host, while in the case of the Samuel et al.\ we choose to show the calculation based on first infall into any halo, since in our accounting we do not separate those with preprocessing that may have accelerated quenching. Note that in the case of Samuel et al., they separately track those satellites that were pre-processed in a low-mass group (LMG) and those that went from being isolated to accreting onto the MW (No-LMG). It is quite clear to see both the satellite-mass dependent pre-processing fraction, and the differences in the quenching times of each satellite type.

It is striking that many trends are quite similar across all three simulations. They all show a transition from mostly quenched to mostly star-forming for satellite \mstar~$\approx 10^8$~\msun. We also see that in our Medium \mkgroup\ bin, the quenched fraction drops below $\sim 50\%$ for \mstar~$\approx 10^8$~\msun. However, we do not observe a universal 2~Gyr quenching time for galaxies with \mstar~$<10^8$~\msun, but rather a continuously increasing $t_q$ with \mstar. In this sense, our results are aligned with a finding of \citeauthor{Akins_2021}, which is that quenching time varies smoothly with satellite mass, because the total gas mass is also rising. 

Overall, our inferred $t_q$ seem systematically longer than are measured directly in the simulations. In the Pan et al.\ work, they find a significant population of promptly quenched satellites ($<1$~Gyr after infall) that dominate the satellites with \mstar$<10^7$~\msun, and then comprises $\sim 50\%$ of the $10^7-10^8$~\msun\ galaxies; our inferred $t_q$ are longer at matching stellar mass. Pan et al.\ also find that their early-type galaxies span a narrower range in NUV color than the ELVES satellites, perhaps due to more rapid quenching leading to a dearth of satellites with more extended star-formation histories. Likewise, the \citeauthor{Samuel_2022} galaxies are accreted into the halo pre-quenched in $>50\%$ of cases even at \mstar~$\approx 10^7$~\msun. 

\citet{Simons:2020} suggests that low-resolution simulations that have a relatively smooth circumgalactic medium cannot capture the proper clumpiness and resulting stochastic nature of ram-pressure stripping \citep[see also][]{Emerick_2016,Hausammann:2019}. The other possibility is that overly strong feedback biases the quenched fraction \citep[e.g.,][]{ElBadry:2018,Kado-Fong:2021aa}. Along those same lines, \citet{Jahn:2022} also looked at LMC-mass halos with FIRE-II simulations. Again, the quenching may be overly efficient, as they report a very similar quenched fraction with satellite mass in the LMC-mass and MW-mass halos. In contrast, we clearly see a lower quenched fraction at all satellite \mstar\ around the Low hosts, with the quenched fraction falling to $\sim 50\%$ at \mstar$\sim 10^7$~\msun, and sitting lower than $f_q$ in MW-mass halos at all \mstar. More statistics for low-mass hosts will help clarify whether our observational results are really in conflict with these simulations \citep[and are on the way][]{Carlin:2021,Garling:2021,Davis:2021,Mutlu-Pakdil:2022}.

\subsection{Non-monolithic quenching}

\begin{figure}
\includegraphics[width=0.45\textwidth]{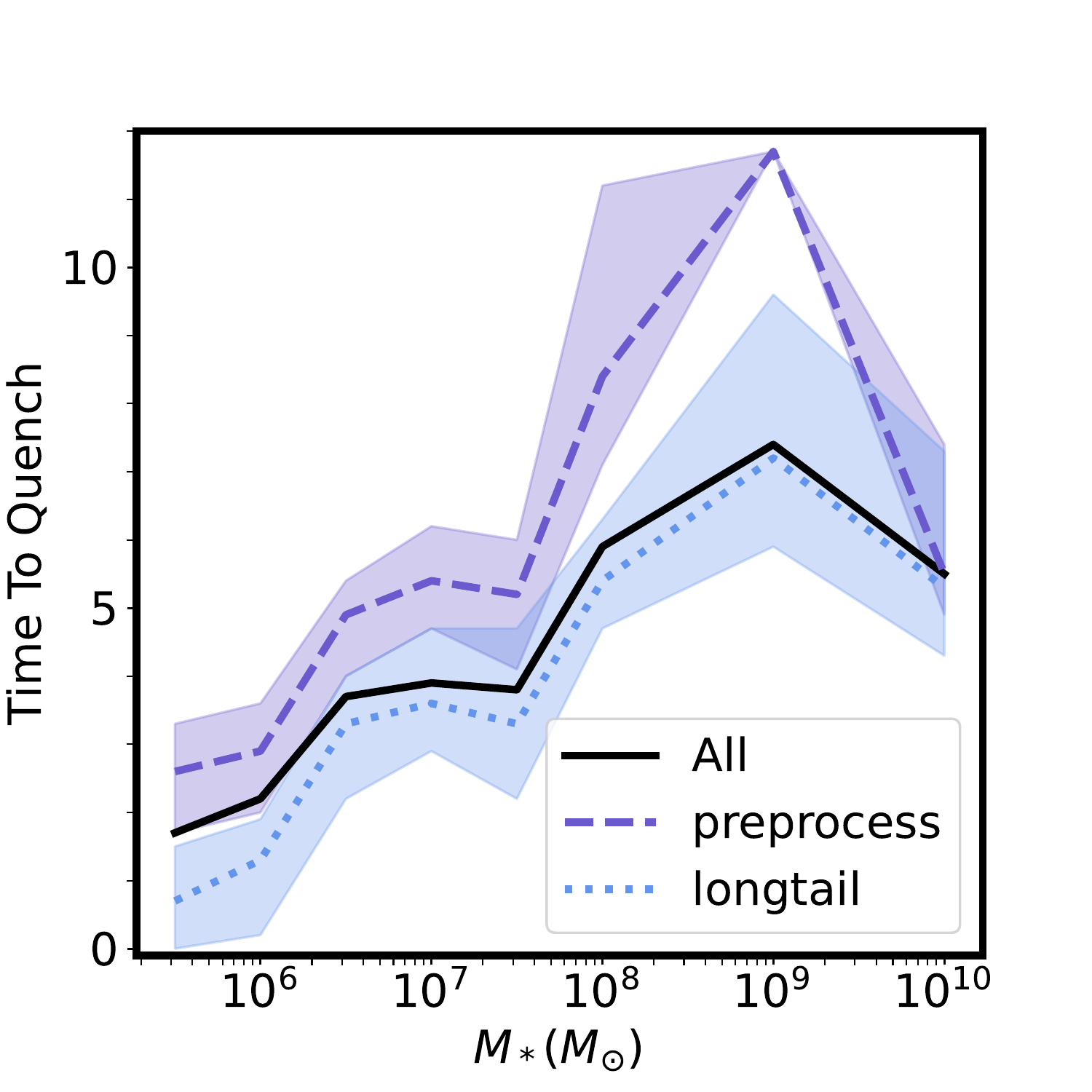}
\caption{Inferred quenching time in satellite stellar mass bins, but exploring the impact of preprocessing or a tail of unquenched things. The preprocessing model assumes that at all masses $M_*<10^8$~\msun, there is a fraction $20\%$ of galaxies that enters the halo quenched. The long tail model assumes quenching acts quickly on timescale $t_q$, but only quenches 80\% of the halos. Then, after 5 Gyr ($M_* < 10^8$~\msun) or 8 Gyr ($M_* < 10^8$~\msun), the quenched fraction goes to 100\%. 
\label{fig:quenchvarymodel}}
\end{figure}

One thing that is clear from comparing with the detailed simulations is that at low satellite mass, $M_* < 10^7$~\msun, many of the satellites are quenched before they accrete into the host. Here, we explore how much longer the quenching time would be for the remaining satellites, if we heuristically account for a pre-quenched fraction. We also crudely explore the idea that there may not be one average quenching time for all satellites.

The first effect that we investigate is that of preprocessing. We know that some satellite galaxies accreted onto the MW from lower-mass groups \citep[e.g. with the Large and Small Magellanic Clouds, ][]{Wetzel_2015} have been in a smaller group environment prior to accreting onto the MW, and may have been environmentally quenched at that time. \citet{Wetzel_2015} use high-resolution cosmological simulations to estimate that at least $\sim 30\%$ of satellites with \mstar$=10^7-10^9$~\msun\ were in another more massive halo before becoming a satellite of the MW \citep[see also][]{Deason_2015,Samuel_2022}. Motivated by this number, we determine an alternate $t_q$ assuming that 30\% of all galaxies with \mstar$<10^{10}$~\msun\ were quenched in a prior halo. The result is shown in Figure \ref{fig:quenchvarymodel} in dashed, where it is clear that positing a 30\% quenched fraction effectively lengthens the inferred quenching time by lowering the number of galaxies that must be quenched. However, the change is not huge; for a 30\% pre-processed fraction we can lengthen the quenching time by $\sim 1$~Gyr.

The second effect we investigate is the idea that there is a distribution of quenching times, and a tail of objects that quench slowly, with the bulk of galaxies still quenching at $t_q$. Again motivated by MW results, we investigate the change to $t_q$ if satellites with \mstar$<10^8$~\msun\ have 20\% of objects quenching in 5~Gyr, while more massive satellites have a tail extending to quenching times of 8~Gyr \citep[e.g.,][]{Akins_2021,Fillingham:2019}. This tweak means that all remaining galaxies must quench more rapidly, to accommodate the 20\% tail of slow-quenching galaxies. Here the difference from our fiducial model is even smaller. 

Of course, the two effects act in opposite directions. Since both must be at play, it is a quantitative question which is more important, and very likely differs from host to host. Furthermore, these two effects are likely to be mass dependent. Progress will require more realistic hydrodynamical simulations \citep[e.g.,][]{Simons:2020}.

\section{Summary}

We look at the quenching properties of a sample of 26 MW-like hosts from the ELVES survey.  With 408 satellites around 26 hosts, of which 313 have secure distances, we study quenched fractions as a function of host stellar and halo mass, host morphology, satellite mass, and satellite-host distance. We find that there is significant spread in the quenched fractions of individual hosts, but no secondary correlation with host halo mass or host stellar mass. Overall, the quenched fraction with satellite stellar mass is similar to that seen in the MW. We see hints that more massive halos are more effective at quenching the most massive satellites, and some marginal evidence that satellites quench more effectively at smaller radial distance. We do not see a difference in quenched fraction between early- and late-type hosts. 

We then infer the average quenching time in bins of satellite stellar mass by combining the observed quenched fractions with a distribution of infall times from the semi-analytic modeling code \satgen. From our lowest to highest mass bin, we find the quenching time steadily rises from 2 to 8 Gyr for $5 \times 10^5-5 \times 10^8$~\msun. Our inferred times are comparable to, but systematically longer than, those found in high-resolution hydrodynamical simulations \citep{Akins_2021,Samuel_2022,Pan:2022}. If ram-pressure stripping is the dominant quenching mechanism, then the stripping must grow less efficient with progressively higher satellite mass. Furthermore, our average quenching times appear to be longer than the quenching delay times reported by recent hydrodynamical simulations. We could underestimate the quenched fractions in the data due to projection and distance errors. At the same time, the simulations may quench galaxies too quickly through feedback prescriptions or averaging over the clumpy circumgalactic medium.

In the future, it would be helpful to measure a robust instantaneous star formation indicator like H$\alpha$ or \ion{H}{1} for a larger fraction of the ELVES galaxies, particularly at the extreme ends of the quenched fraction range. It would be even more informative to have detailed star formation histories for the galaxies based on resolved color-magnitude diagrams; thus far, this has only been done within 4 Mpc \citep[e.g.,][]{Weisz_2011}, but with the \emph{James Webb Space Telescope} and then the \emph{Nancy Grace Roman Space Telescope} \citep{Akeson:2019}, there is hope to use luminous infrared populations to probe quenching times to larger distance \citep[e.g.,][]{Melbourne_2012}.

\section*{Acknowledgment}
We acknowledge H.\ Akins, J.\ Samuel, C.\ Simpson, and A.\ Wetzel for kindly sharing their data. We thank M.\ Putman, Y. Mao, and J.\ Zhu for helpful discussions that improved this manuscript. We thank the referee for a timely and constructive report that significantly improved this work. JEG gratefully acknowledges support from NSF grant AST-2106730.
S.D. is supported by NASA through Hubble Fellowship grant HST-HF2-51454.001-A awarded by the Space Telescope Science Institute, which is operated by the Association of Universities for Research in Astronomy, Incorporated, under NASA contract NAS5-26555.

ELVES is based in part on observations obtained with MegaPrime/MegaCam, a joint project of CFHT and CEA/DAPNIA, at the Canada-France-Hawaii Telescope (CFHT) which is operated by the National Research Council (NRC) of Canada, the Institut National des Science de l'Univers of the Centre National de la Recherche Scientifique (CNRS) of France, and the University of Hawaii. The observations at the Canada-France-Hawaii Telescope were performed with care and respect from the summit of Maunakea which is a significant cultural and historic site. 

\newpage

\section{Appendix}

\begin{figure}
\includegraphics[width=0.45\textwidth]{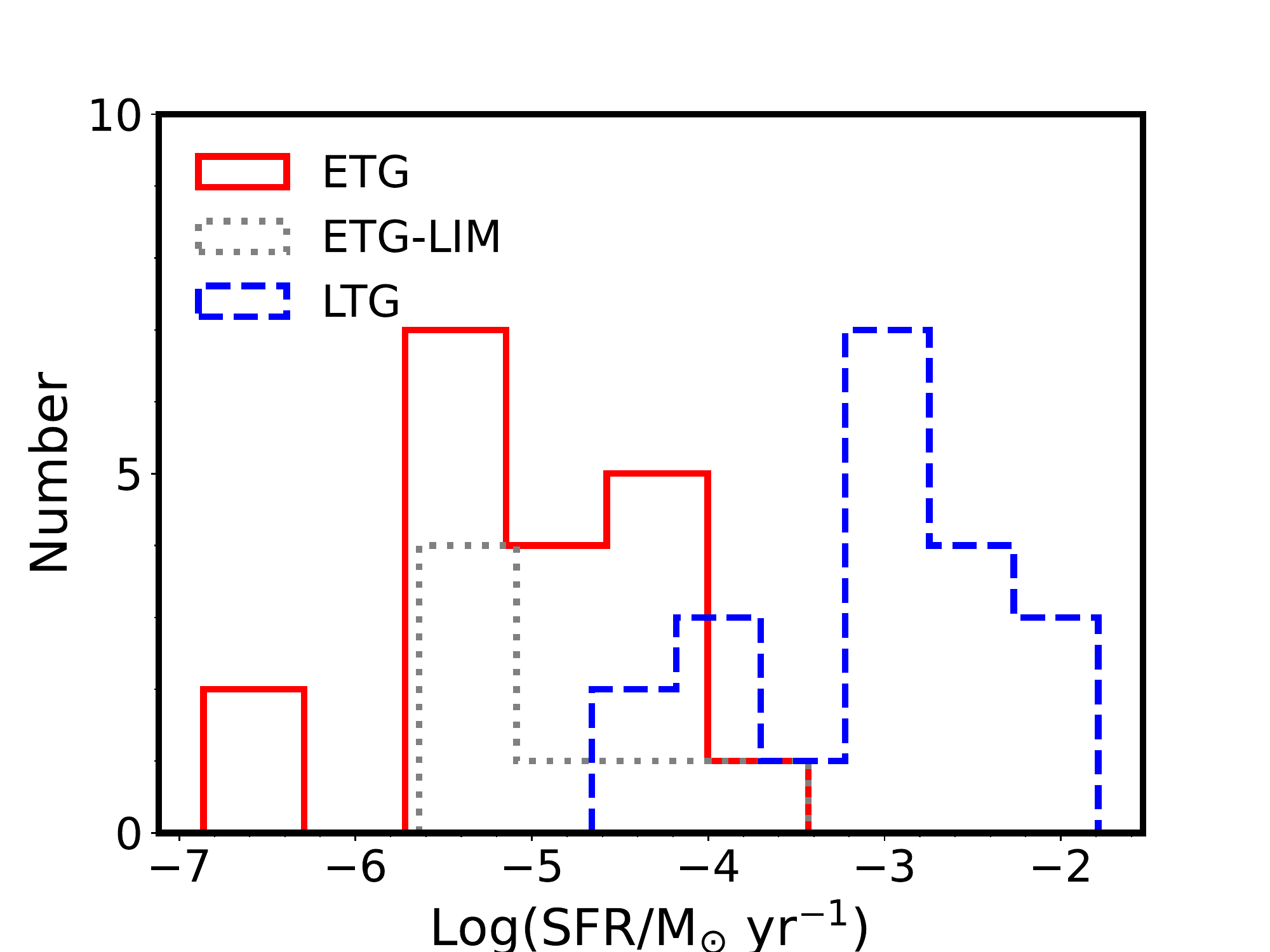}
\caption{Histogram of star-formation rates inferred from late-type and early-type satellites within ELVES in the \mstar$=10^7-10^8$~\msun\ range. We show detections in early and late-type galaxies, as well as the reported upper-limits (ETG-LIM) in early-type galaxies (all late-types are detected). There is a clear bi-modality in which the morphologically flagged early-type galaxies have either much lower star-formation rates or their H$\alpha$ arises from different physical processes. 
\label{fig:sfHa}}
\end{figure}

Here we present the inferred star-formation rates for the ELVES targets with cross-matches in the \citet{Kaisin:2019} catalog \citep[see also][]{Karachentsev:2021}. The H$\alpha$ fluxes are based on narrow-band imaging, and generally have a depth of $\sim 5 \times 10^{-14}$~erg~s~cm$^2$ or EWs of a few \AA. We use the relation between H$\alpha$ luminosity and star-formation rate from \citet{Kennicutt:2012aa} to calculate the star-formation rate assuming that all H$\alpha$ arises from star formation \citep[which need not be true; e.g.,][]{Yan:2018,Belfiore:2022}. The result is shown in Figure \ref{fig:sfHa}. 

\bibliographystyle{aasjournal}
\bibliography{Greco_et_al_2018.bib}

\end{document}